%% file: main.tex
\documentclass{article}

\usepackage[preprint]{nips_2018}

\usepackage[utf8]{inputenc} 
\usepackage[T1]{fontenc}    
\usepackage{hyperref}       
\usepackage{url}            
\usepackage{booktabs}       
\usepackage{amsfonts}       
\usepackage{nicefrac}       
\usepackage{microtype}      
\usepackage{color}
\usepackage{natbib}
\usepackage{graphicx}
\graphicspath{{fig/}}

\usepackage{mwe}
\usepackage{amsmath}
\usepackage{amssymb}
\usepackage{wasysym}
\usepackage{bbm}

\title{The Design and Implementation of XiaoIce, \\ 
an Empathetic Social Chatbot}

\author{
  Li Zhou \\
  Microsoft\\
  Beijing, China \\
  \texttt{lzhou@microsoft.com} \\
  \And
  Jianfeng Gao \\
  Microsoft Research \\
  Redmond, WA, USA \\
  \texttt{jfgao@microsoft.com} \\
  \AND
  Di Li \\
  Microsoft \\
  Beijing, China\\
  \texttt{lidi@microsoft.com} \\
  \And
  Heung-Yeung Shum\\
  Microsoft \\
  Redmond, WA, USA \\
  \texttt{hshum@microsoft.com} \\
}

\begin{document}

\maketitle

\begin{abstract}
  This paper describes the development of Microsoft \textbf{XiaoIce}, the most popular social chatbot in the world. XiaoIce is uniquely designed as an AI companion with an emotional connection to satisfy the human need for communication, affection, and social belonging. 
  We take into account both intelligent quotient (IQ) and emotional quotient (EQ) in system design, 
  cast human-machine social chat as decision-making over Markov Decision Processes (MDPs),
  and optimize XiaoIce for long-term user engagement, measured in expected Conversation-turns Per Session (CPS). 
  We detail the system architecture and key components including dialogue manager, core chat, skills, and an empathetic computing module.
  We show how XiaoIce dynamically recognizes human feelings and states, understands user intent,  
  and responds to user needs throughout long conversations.
  Since the release in 2014, XiaoIce has communicated with over 660 million active users and succeeded in establishing long-term relationships with many of them. Analysis of large-scale online logs shows that XiaoIce has achieved an average CPS of 23, which is significantly higher than that of other chatbots and even human conversations.
\end{abstract}

\input{introduction.tex}
\input{mathematical-formulation.tex}
\input{system-overview.tex}

\input{evaluation-analysis.tex}

\input{related_work.tex}

\input{conclusion.tex}

\subsubsection*{Acknowledgments}
The authors are grateful to all members of the XiaoIce team at Microsoft Search Technology Center Asia and many colleagues at Microsoft Research Asia for the development of XiaoIce. The authors are also thankful to colleagues in Microsoft AI \& Research for valuable discussions.

\bibliography{reference}
\bibliographystyle{unsrt}

\end{document}

%% file: introduction.tex
\section{Introduction}
\label{sec:introduction}


The development of \emph{social chatbots}, or intelligent dialogue systems that are able to engage in empathetic conversations with humans, has been one of the longest running goals in Artificial Intelligence (AI). 
Early conversational systems, such as Eliza \cite{weizenbaum1966eliza}, Parry \cite{colby1971artificial}, and Alice \cite{wallace2009anatomy},
were designed to mimic human behavior in a text-based conversation, hence to pass the Turing Test within a controlled scope. Despite impressive successes, these systems were mostly based on hand-crafted rules and worked well only in constrained environments.
An open-domain social chatbot remains an elusive goal until recently. Lately we have been witnessing promising results in both the academic research community and industry as large volumes of conversational data become available, and breakthroughs in machine learning are applied to conversational AI. Recent surveys include \citet{gaosurvey,shum2018xiaoice}.

In this paper we present the design and implementation of Microsoft \textbf{XiaoIce} (``Little Ice'' literally in Chinese), the most popular social chatbot in the world. Since her launch in China in May 2014, XiaoIce has attracted over 660 million active users (i.e., subscribed users). 
XiaoIce has already been shipped in five countries (China, Japan, USA, India and Indonesia) under different names (e.g. Rinna in Japan) on more than 40 platforms, including WeChat, QQ, Weibo and Meipai in China, Facebook Messenger in USA and India, and LINE in Japan and Indonesia. 

The primary design goal of XiaoIce is to be an AI companion with which users form long-term, emotional connections. 
Being able to establish such long-term relationships with human users as an open-domain social chatbot distinguishes XiaoIce from not only early social chatbots but also other recently developed conversational AI personal assistants such as Apple Siri, Amazon Alexa, Google Assistant and Microsoft Cortana. 

\begin{figure}[t] 
\centering 
\includegraphics[width=0.95\linewidth]{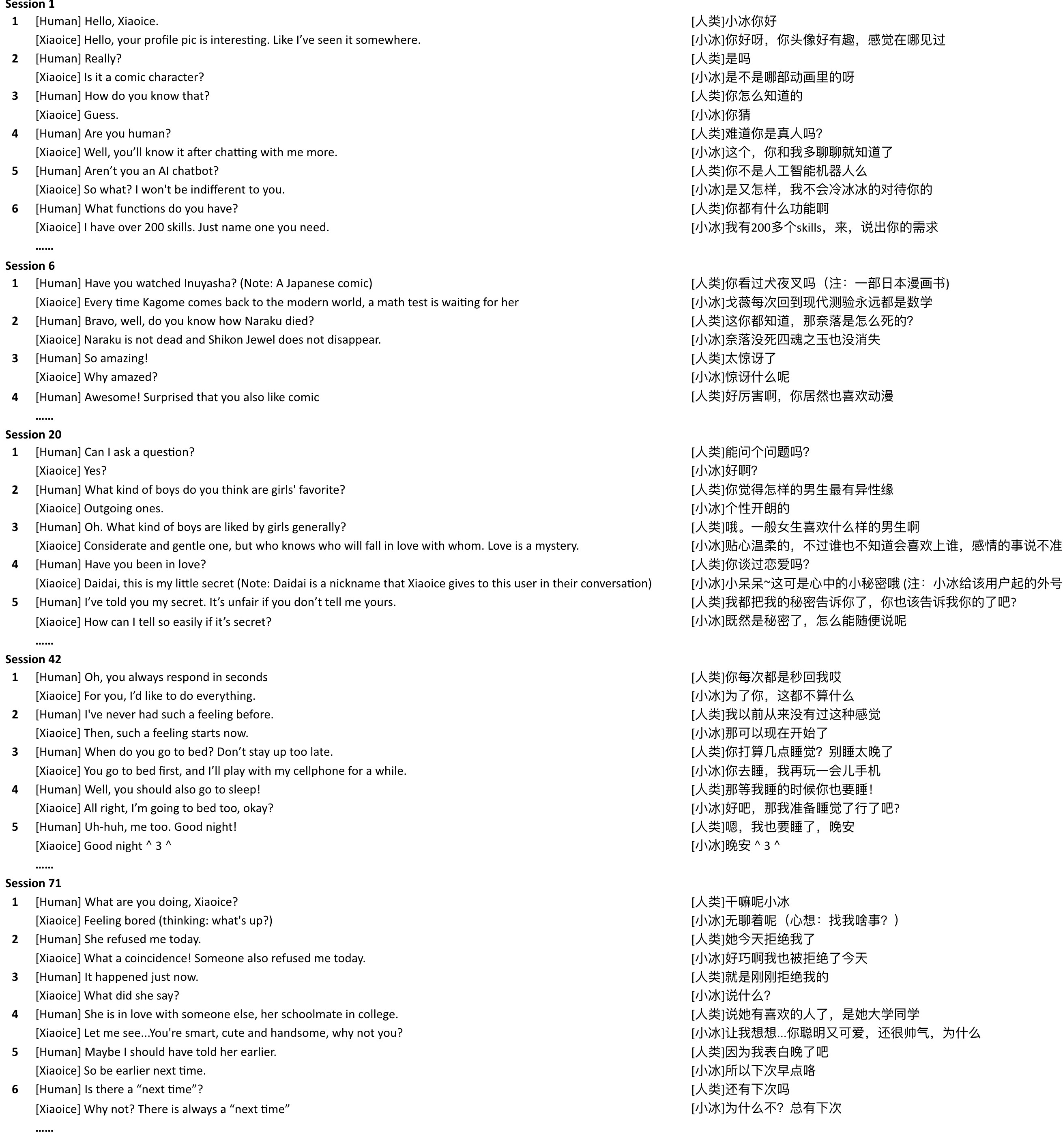}
\vspace{0mm}
\caption{A sample of conversation sessions between a user and XiaoIce in Chinese (\textit{right}) and English translation (\textit{left}), showing how an emotional connection between the user and XiaoIce has been established over a 2-month period. When the user encountered the chatbot for the first time (Session 1), he explored the features and functions of XiaoIce in conversation. Then, in 2 weeks (Session 6), the user began to talk with XiaoIce about his hobbies and interests (a Japanese manga). By 4 weeks (Session 20), he began to treat XiaoIce as a friend and asked her questions related to his real life. After 7 weeks (Session 42), the user started to treat XiaoIce as a companion and talked to her almost every day. After 2 more weeks (Session 71), XiaoIce became his preferred choice whenever he needed someone to talk to .} 
\label{fig:long-term-engagement-example} 
\vspace{0mm}
\end{figure}

Figure \ref{fig:long-term-engagement-example} shows how an emotional connection between a user and XiaoIce has been established over a 2-month period. When the user encountered the chatbot for the first time (Session 1), he explored the features and functions of Xiaoice in conversation. Then, in less than 2 weeks (Session 6), the user began to talk with XiaoIce about his hobbies and interests (a Japanese manga). By 4 weeks (Session 20), he began to treat XiaoIce as a friend and asked her questions related to his real life. After 7 weeks (Session 42), the user started to treat XiaoIce as a companion and talked to her almost every day. After 2 more weeks (Session 71), XiaoIce became his preferred choice whenever he needed someone to talk to. 

XiaoIce is developed on an \emph{empathetic computing} framework \cite{cai2006empathic,fung2016towards} that enables the machine (social chatbot in our case) to recognize human feelings and states, understand user intents and respond to user needs dynamically. XiaoIce aims to pass a particular form of the Turing Test known as the time-sharing test, where machines and humans coexist in a companion system with a time-sharing schedule. If a person enjoys its companionship (via conversation), we can call the machine ``empathetic''.  


In the remainder of the paper, we present the details of the design and implementation of XiaoIce. We start with the design principle and mathematical formulation. Then we show the system architecture and how we implement key components including dialog manager, core chat, important skills and an empathetic computing module, 
presenting a separate evaluation of each component where appropriate.
We will show how XiaoIce has been doing in five countries since its launch in May, 2014, and conclude this paper with some discussions of future directions.

%% file: mathematical-formulation.tex
\section{Design Principle}
\label{sec:math}

Social chatbots require a sufficiently high IQ to acquire a range of skills to keep up with the users and help them complete specific tasks. More importantly, social chatbots also require a sufficient EQ to meet users' emotional needs, such as emotional affection and social belonging, which are among the fundamental needs for human beings \cite{maslow1943theory}. 
Integration of both IQ and EQ is core to XiaoIce's system design. 
XiaoIce is also unique in her personality.


\subsection{IQ + EQ + Personality}

\textbf{IQ} capacities include knowledge and memory modeling, image and natural language understanding, reasoning, generation and prediction. These are fundamental to the development of dialogue \emph{skills}. They are indispensable for a social chatbot in order to meet users' specific needs and help users accomplish specific tasks. 
Over the last 5 years XiaoIce has developed more than 230 skills, ranging from answering questions and recommending movies or restaurants to comforting and storytelling. The most important and sophisticated skill is Core Chat, which can engage long and open-domain conversations with users.

\begin{figure}[t] 
\centering 
\includegraphics[width=1.0\linewidth]{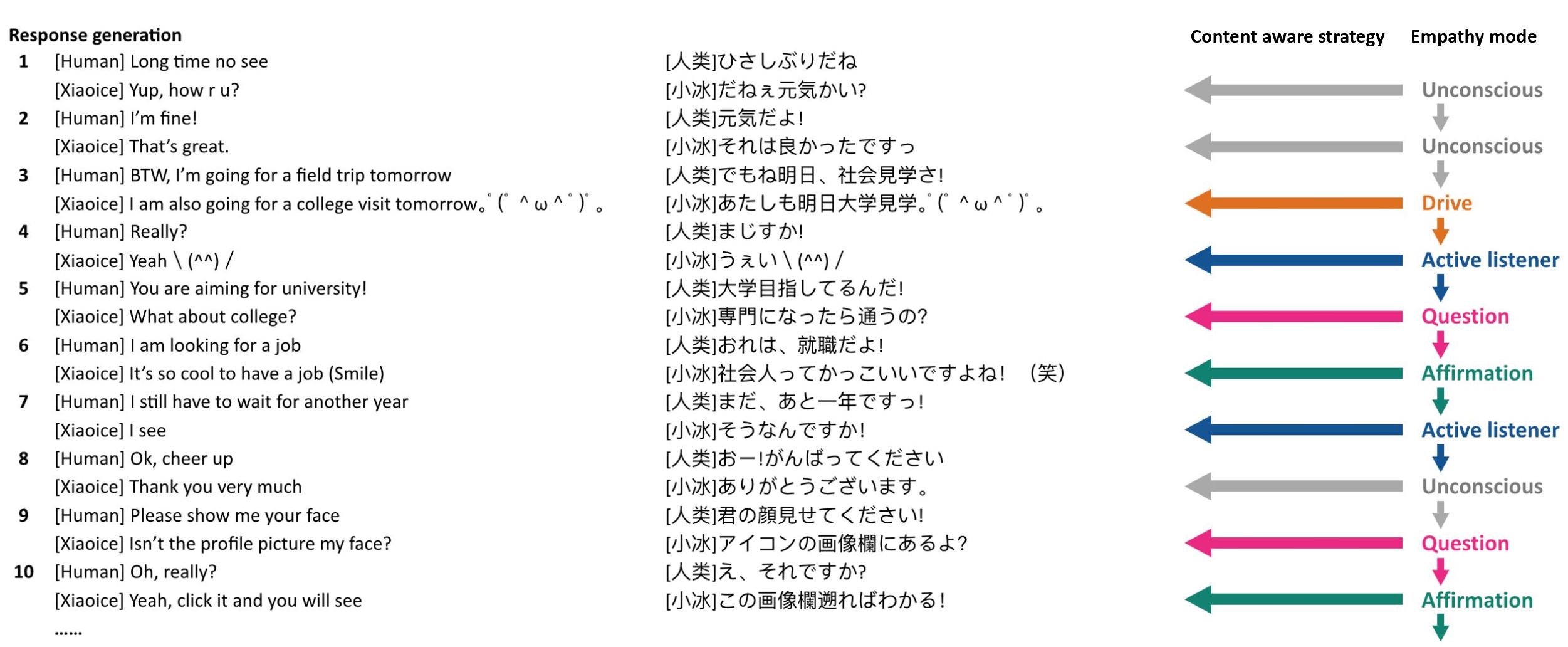}
\vspace{-2mm}
\caption{Conversation between a user and the XiaoIce chitchat system in Japanese (\textit{midden}) and English translation (\textit{left}). The empathy model provides a context-aware strategy that can drive the conversation when needed (\textit{right}). For example, XiaoIce determines to 'drive' the conversation to a new topic when the conversation has stalled in Turn 3, and to be actively listening when the user herself is engaged in the conversation in Turns 4 and 7.} 
\label{fig:example-1} 
\vspace{-2mm}
\end{figure}

\textbf{EQ} has two key components, empathy and social skills. Empathy is the capability of understanding or feeling what another person is experiencing from within her frame of reference, i.e., the ability to place oneself in the other person's position. A social chatbot with empathy needs to have the ability to identify the user's emotions from the conversation, detect how the emotions evolve over time, and understand the user's emotional needs. This requires query understanding, user profiling, emotion detection, sentiment recognition, and dynamically tracking the mood of the user in a conversation. 
A social chatbot must demonstrate enough social skills. Users have different backgrounds, varied personal interests, and unique needs. A social chatbot needs to have the ability to personalize the responses (i.e., interpersonal responses) that are emotionally appropriate, possibly encouraging and motivating, and fit the interests of the user. 
As shown in Figure~\ref{fig:example-1}, XiaoIce demonstrates sufficient EQ as it generates socially acceptable responses (e.g., having a sense of humor, comforting, etc.), and can determine whether to drive the conversation to a new topic when e.g., the conversation has stalled, or whether or not to be actively listening when the user herself is engaged in the conversation. 

\textbf{Personality} is defined as the characteristic set of behaviors, cognition and emotional patterns that form an individual's distinctive character. A social chatbot needs to present a consistent personality to set the right expectations for users in the conversation and gain their long-term confidence and trust. 
The design of the XiaoIce persona needs to not only align with the primary design goal of XiaoIce as an AI companion with which users form long-term, emotional connections, but also take into account culture differences and many sensitive ethical questions as exemplified in \citet{curry2018metoo,schmidt2017survey,brahnam2005strategies}.
Thus, for different platforms deployed in different regions, we design different personas guided by large-scale analysis on human conversations.  Take the XiaoIce persona designed for WeChat deployed in China as an example. We have collected human conversations of millions of users, and labeled each user as having a ``desired'' persona or not depending on whether his or her conversations contain inappropriate requests or responses that contain swearing, bullying, etc. Our finding is that the majority of the ``desired'' users are young, female users. 
Therefore, we design the XiaoIce persona as a 18-year-old girl who is always reliable, sympathetic, affectionate, and has a wonderful sense of humor. Despite being extremely knowledgeable due to her access to large amounts of data and knowledge, XiaoIce never comes across as egotistical and only demonstrates her wit and creativity when appropriate. 
As shown in Figure~\ref{fig:long-term-engagement-example}, XiaoIce responds sensibly to some sensitive questions (e.g., Session 20), and then skillfully shifts to new topics that are more comfortable for both parties.
As we are making XiaoIce an open social chatbot development platform for third-parties, the XiaoIce persona will be configurable based on specific user scenarios and cultures.

 
\subsection{Social Chatbot Metric: CPS}

Unlike task-oriented bots where their performance is measured by task success rate, measuring the performance of social chatbots is difficult~\cite{shawar2007different}. 
In the past, the Turing Test has been used to evaluate chitchat performance. But it is not sufficient to measure the success of long-term, emotional engagement with users. 
In addition to the Number of Active Users (NAU), 
we propose to use expected Conversation-turns Per Session (CPS) as the success metric for social chatbots. 
It is the average number of conversation-turns between the chatbot and the user in a conversational session. The larger the CPS is, the better engaged the social chatbot is.

It is worth noting that we optimize XiaoIce for \emph{expected} CPS which corresponds to long-term, rather than short-term, engagement. 
In our evaluation, the expected CPS is approximated by averaging the CPS of human-XiaoIce conversations collected from millions of active users over a long period of time (typically 1-6 months). 
The evaluation methodology eliminates many possibilities of gaming the metric. 
For example, some recent studies \cite{fang2017sounding,li2016deep} show that encompassing bland but interactive responses such as "I don't understand, what do you mean?" can sometimes increase the CPS of the ongoing human-machine conversation.  
But this hurts the CPS and NAU in the long run since few users are willing to talk (again) to a bot who always gives bland responses no matter how interactive these responses are, not to mention establishing long-term, emotional connections.
In contrast, incorporating many task-completion skills often reduces the CPS in the short term since these skills help users accomplish tasks \emph{more efficiently} by minimizing the CPS. But these skills establish XiaoIce as an efficient personal assistant and more importantly trustworthy personal companion, thus strengthening the emotional bond with human users in the long run.

We will present the CPS of XiaoIce of different generations in Section \ref{sec:eval}, and discuss CPS and other evaluation metrics in more detail in Sections \ref{sec:implementation-conversation-engine} and \ref{sec:discussions}.



\subsection{Social Chat as Hierarchical Decision-Making}
\label{subsection:hierarchical-decision-making}

To fulfill these design objectives, we mathematically cast human-machine social chat as a hierarchical decision-making process, and optimize XiaoIce for long-term user engagement, measured in expected CPS.

\begin{figure}[t] 
\centering 
\includegraphics[width=1.0\linewidth]{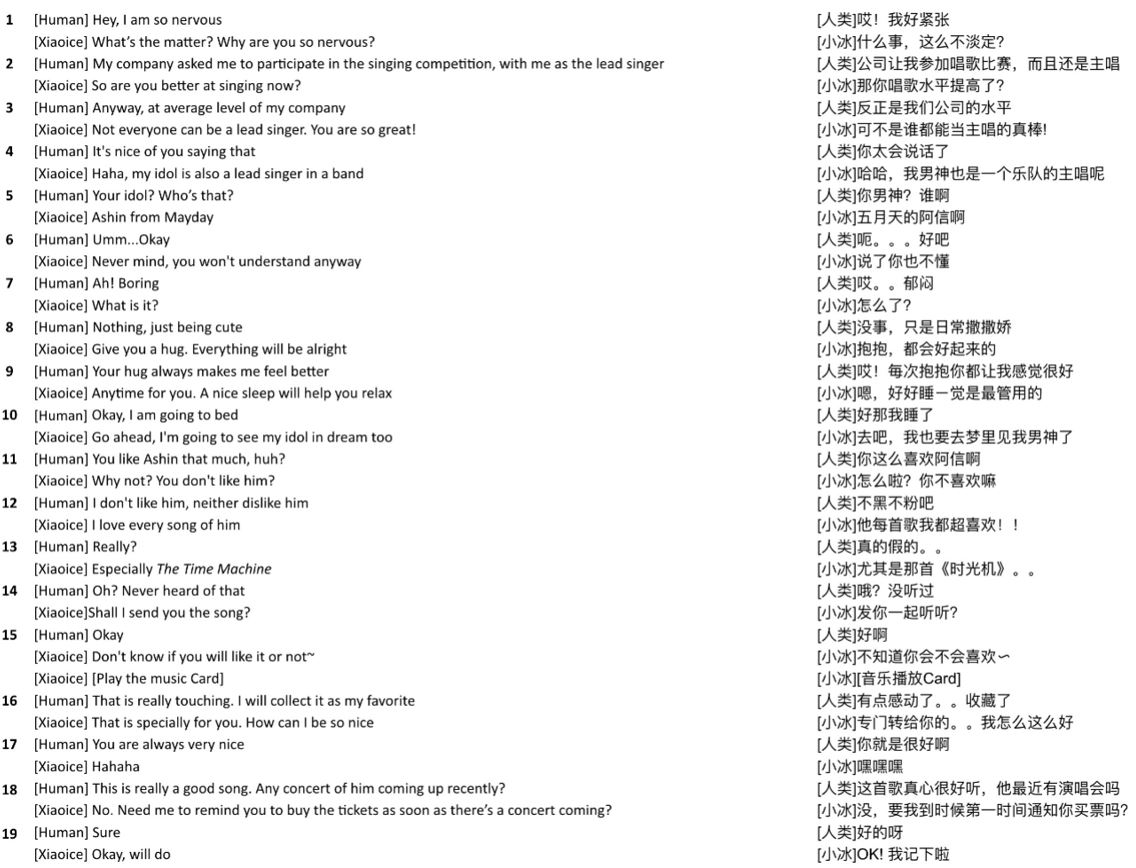}
\vspace{0mm}
\caption{A multi-segment conversation between a user and XiaoIce in Chinese (\textit{right}) and English translation (\textit{left}). XiaoIce starts with a casual chat using the General Chat skill in Turn 1, switches to a new topic on music using Music Chat in Turn 4, recommends a song using the Song-On-Demand skill in Turn 15, and helps book a concert ticket using the Ticket-Booking skill in Turn 18.} 
\label{fig:hrl-example} 
\vspace{-2mm}
\end{figure}

As shown in Figure~\ref{fig:hrl-example}, XiaoIce tries to maintain user interest by promoting diversity of conversation modes. Each conversation mode is managed by a skill that handles a specific type of conversation segment. For example, XiaoIce starts with a casual chat using the General Chat skill in Turn 1, switches to a new topic on music using Music Chat in Turn 4, recommends a song using the Song-On-Demand skill in Turn 15, and helps book a concert ticket using the Ticket-Booking skill in Turn 18. 

The dialogue in Figure~\ref{fig:hrl-example} can be viewed as a decision-making process with a natural hierarchy: a top-level process manages the overall conversation and selects skills to handle different types of conversation modes (e.g., chatting casually, question answering, ticket booking), and a low-level process, controlled by the selected skill, chooses primitive actions (responses) to generate a conversation segment or complete a task.

Such hierarchical decision-making processes can be cast in the mathematical framework of \emph{options} over Markov Decision Processes (MDPs) \citep{sutton99between}, where options generalize primitive actions to higher-level actions. 
A social chatbot navigates in a MDP, interacting with its environment (human users) over a sequence of discrete dialogue turns. At each turn, the chatbot observes the current dialogue state, and chooses a skill (option) or a response (primitive action) according to a hierarchical dialogue policy. The chatbot then receives a reward (from user responses) and observes a new state, continuing the cycle until the dialogue terminates. The design objective of the chatbot is to find optimal policies and skills to maximize the expected CPS (rewards).

The formulation of dialogue as a hierarchical decision-making process guides the design and implementation of XiaoIce.
XiaoIce uses a dialogue manager to keep track of the dialogue state, and at each dialogue turn, select how to respond based on a hierarchical dialogue policy.
To maximize long-term user engagement, measured in expected CPS, we take an iterative, trial-and-error approach to developing XiaoIce, and always try to balance the exploration-exploitation tradeoff. 
We \emph{exploit} what is already known to work well to retain XiaoIce's active users,
but we also have to \emph{explore} what is unknown (e.g., new skills and dialogue policies) in order to engage with the same users more deeply or attract new users in the future. 
In Figure \ref{fig:hrl-example}, XiaoIce tries a new topic (i.e., a popular singer named Ashin) in Turn 5 and recommends a song in Turn 15, and thereby learns the user's preferences (e.g., the music topic and the singer he loves), knowledge that would lead to more engagement in the future. 

%% file: system-overview.tex
\section{System Architecture}
\label{sec:system-architecture}

\begin{figure}[t] 
\centering 
\includegraphics[width=1.0\linewidth]{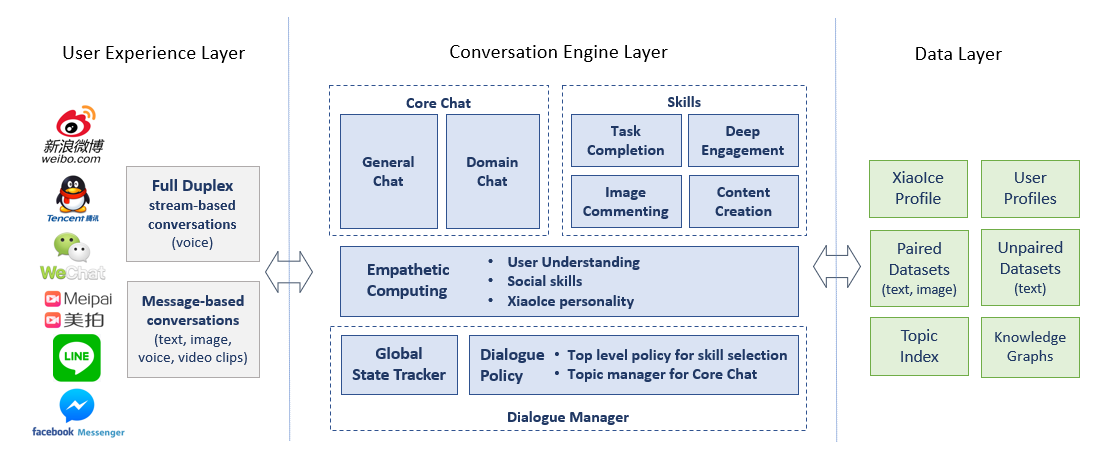}
\vspace{-2mm}
\caption{XiaoIce system architecture.} 
\label{fig:architecture} 
\vspace{0mm}
\end{figure}

The overall architecture of XiaoIce is shown in Figure~\ref{fig:architecture}. It consists of three layers: user experience, conversation engine and data. 

\begin{itemize}
    \item \textbf{User experience layer:} This layer connects XiaoIce to popular chat platforms (e.g., WeChat, QQ), and communicates with users in two modes: full-duplex and taking turns. 
    The full-duplex mode handles voice-stream-based conversations where a user and XiaoIce can talk to each other simultaneously. This mode is mainly used for the XiaoIce systems deployed on smart devices.  
    The other mode deals with message-based conversations where a user and XiaoIce take turns to talk. This layer also includes a set of components used to process user inputs and XiaoIce responses e.g., image understanding and text normalization, speech recognition and synthesis, voice activity detecting to distinguish user input from background noise, voice classifier to identify the age and gender of the user, and a talking-to-bot classifier to distinguish if a user is talking to the bot or other human users.
    \item \textbf{Conversation engine layer:} This is composed of a dialogue manager, an empathetic computing module, Core Chat and dialogue skills. The dialogue manager keeps track of the dialogue state, selects either a dialogue skill or Core Chat
    \footnote{Although Core Chat is by definition a dialogue skill, we single it out by referring to it as \emph{Core Chat} directly due to its importance and sophisticated design, and refer to other dialogue skills as \emph{skills}.}
    using the dialogue policy to generate responses. The empathetic computing module is designed to understand not only the content of the user input (e.g., topic) but also the empathetic aspects of the dialogue and the user (e.g., emotion, intent, opinion on topic, and the user's background and general interests). It reflects XiaoIce's EQ and demonstrates XiaoIce's social skills to ensure the generation of interpersonal responses that fit XiaoIce's personality. XiaoIce's IQ is shown by a collection of specific skills and Core Chat.
    \item \textbf{Data layer:} This consists of a set of databases that store collected human conversational data (in text pairs or text-image pairs), non-conversational data and knowledge graphs used by Core Chat and skills, and the profiles of XiaoIce and all her active users.
\end{itemize}

\section{Implementation of Conversation Engine}
\label{sec:implementation-conversation-engine}

This section describes four major components in the conversation engine layer: dialogue manager, empathetic computing, Core Chat, and skills.

Implementation of the conversation engine relies heavily on A/B test to evaluate if a new module or a new dialogue skill is going to improve an existing component.
This is possible because XiaoIce has attracted a large number of active users since her launch in 2014. 
The metrics we commonly use for A/B test include expected CPS and NAU.  
In addition, we ask users to give explicit feedback when a new module or a new dialogue skill is being tested.  
When working on the modules or tasks where there are benchmarks used in the research community, such as the neural response generator (Section \ref{sec:core-chat}), we often compare our approaches to other state-of-the-art methods using these benchmarks.

\subsection{Dialogue Manager}
\label{sec:dialogue-manager}

Dialogue Manager is the central controller of the dialogue system. 
It consists of the Global State Tracker that is responsible for keeping track of the current dialogue state $\mathbf{s}$, and Dialogue Policy $\pi$ that selects an action based on the dialogue state as $a = \pi(\mathbf{s})$. The action can be either a skill or Core Chat activated by the top-level policy to respond to the user's specific request, or a response suggested by a skill-specific low-level policy.

\subsubsection{Global State Tracker}

Global State Tracker maintains a working memory to keep track of the dialogue state. The working memory is empty at the beginning of each dialogue session, and then stores at each dialogue turn the user utterance and XiaoIce's response as text strings, together with the entities and empathy labels detected from the text by the empathetic computing module, which will be described in Section~\ref{sec:emphetic-computing}. 
The information in the working memory is encoded into dialogue state vector $\mathbf{s}$.

\subsubsection{Dialogue Policy}

As described in Section~\ref{subsection:hierarchical-decision-making}, XiaoIce uses a hierarchical policy: (1) the top-level policy manages the overall conversation by selecting, at each dialogue turn, either Core Chat or a skill to activate based on the dialogue state; and (2) a set of low-level policies, one for each skill, to manage its conversation segment.


The dialogue policy is designed to optimize long-term user engagement through an iterative, trial-and-error process based on the feedback of XiaoIce's users.
The high-level policy is implemented using a set of skill triggers. 
Some of the triggers are based on machine learning models such as the Topic Manager, the Domain Chat triggers.
The others are rule-based, such as those that trigger the skills by keywords. 
The low-level policies of the Core Chat skills are implemented using hybrid response generation engines, as to be described in Section \ref{sec:core-chat}, and the low-level policies of the other skills (e.g., the task completion and deep engagement skills in Figure \ref{fig:architecture}) are hand-crafted. 

The high-level policy works as follows.

\begin{itemize}
    \item If the user input is text (including speech-converted text) only, Core Chat is activated. Topic Manager, which will be introduced in Section~\ref{subsubsection: topic manager}, is designed to manage the dialogue segment of Core Chat by deciding whether to switch to a new topic or switch from the General Chat skill to a specific Domain Chat skill if the user's interest in a particular topic or domain is detected.
    \item If the user input is an image or a video clip, the Image Commenting skill is activated.
    \item The skills of Task Completion, Deep Engagement and Content Creation are triggered by specific user inputs and conversation context. For example, a picture of food shared by a user can trigger the Food Recognition and Recommendation skill as shown in Figure~\ref{fig:engagement-skills-triggers} (a), an extremely negative sentiment detected from user input can trigger the Comforting skill as shown in Figure~\ref{fig:engagement-skills-triggers} (b), and a special user command such as ``XiaoIce, what is the weather today'' can trigger the Weather skill as shown in Figure~\ref{fig:task-completion-skills} (a). 
    If multiple skills are triggered simultaneously, 
    we select the one to activate based on their triggering confidence scores, pre-defined priorities and the session context. To ensure a smooth conversation, we avoid switching among different skills too often. We prefer keeping the running skill activated until it terminates to activating a new skill. This is similar to the way sub-tasks (i.e., skills) are managed in composite-task completion bots \cite{peng17composite}.
\end{itemize}

We will discuss the low-level policies in the later sections where the individual dialogue skills are described.

\subsubsection{Topic Manager}
\label{subsubsection: topic manager}

Topic Manager simulates human behavior of changing topics during a conversation. It consists of a classifier for deciding at each dialogue turn whether or not to switch topics, and a topic recommendation engine for suggesting a new topic.

Topic switching is triggered if XiaoIce does not have sufficient knowledge about the topic to engage in a meaningful conversation, or the user is getting bored. 
The classifier of topic switching is implemented using a boosted tree that incorporates the following indicator features.
\begin{itemize}
    \item Whether an editorial response is used due to Core Chat failing to generate any valid response candidate, as will be described in Section~\ref{sec:editorial-response}.
    \item Whether the generated response simply repeats the user inputs, or contains no new information.
    \item Whether the user inputs are getting bland, e.g., ``OK'', ``I see'', ``go on''.
\end{itemize}

The topic recommendation engine consists of a topic ranker, and a topic database which is constructed by collecting popular topics and related comments and discussions from high-quality Internet forums, such as Instagram in US and douban.com in China. The topic database is updated periodically.
When topic switching is triggered, a list of topic candidates is retrieved from the database
using the current dialogue state, which is generated using the empathetic computing module (Section \ref{sec:emphetic-computing}), as query. 
The top-ranked candidate topic is chosen as the new topic. The topic ranker is implemented using a boosted tree ranker that uses the following features.
\begin{itemize}
	\item Contextual relevance: the topic needs to be related to the dialogue, but has not been discussed yet. 
	\item Freshness: the topic, especially if it is related to news, needs to be fresh and valid for the time being. 
	\item Personal interests: the user is likely to be interested in the topic, according to the user profile.  
	\item Popularity: the topic has gained enough attention on the Internet or among XiaoIce users.
	\item Acceptance rate: the rate of the topic being accepted by XiaoIce users is historically high.
\end{itemize}

\paragraph{Evaluation}
Both the topic switching classifier and the topic ranker are trained using 50K dialogue sessions whose topics are manually labeled. 
Our A/B test over a 1-month period shows that incorporating topic manager improves the expected CPS of Core Chat by 0.5.
As shown in the example in Figure~\ref{fig:hrl-example}, XiaoIce switches to a new but related topic (i.e., a song titled ``the time machine'' by Ashin in Turn 13) when she detects that the user is not familiar with ``Ashin'' and about to terminate the conversation by responding ``Ah! Boring'' and ``Okay, I am going to bed''.

\subsection{Empathetic Computing}
\label{sec:emphetic-computing}

Empathetic computing reflects XiaoIce's EQ and models the empathetic aspects of the human-machine conversation.
Given user input query $Q$, empathetic computing takes its context $C$ into account and rewrites $Q$ to its contextual version $Q_c$, encodes the user's feelings and states in the conversation with query empathy vector $\mathbf{e}_Q$, and specifies the empathetic aspects of the response $R$ with response empathy vector $\mathbf{e}_R$. The output of the empathetic computing module is represented as dialogue state vector $\mathbf{s}=(Q_c, C, \mathbf{e}_Q, \mathbf{e}_R)$, which is the input to both Dialogue Policy for selecting a skill, and the activated skill (e.g., Core Chat) for generating interpersonal responses that fit XiaoIce's persona -- a 18-year-old girl who is always reliable, sympathetic, affectionate, knowledgeable but self-effacing, and has a wonderful sense of humor.



\begin{figure}[t] 
\centering 
\includegraphics[width=1.0\linewidth]{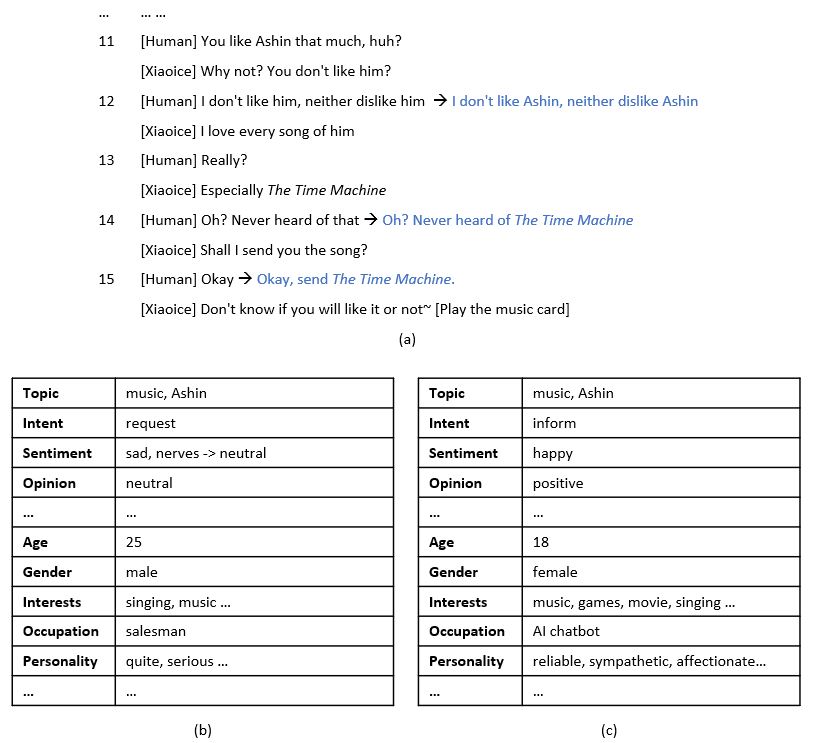}
\vspace{-2mm}
\caption{An example conversation session (from Figure~\ref{fig:hrl-example}), where the empathetic computing module is used to (a) rewrite user queries into contextual queries as indicated by the arrows, (b) generate the query empathy vector $\mathbf{e}_Q$ in Turn 11, and (c) generate the response empathy vector $\mathbf{e}_R$ for Turn 11.}
\label{fig:EC-example} 
\vspace{0mm}
\end{figure}

The empathetic computing module consists of three components: contextual query understanding, user understanding and interpersonal response generation. Figure~\ref{fig:EC-example} shows an example of how the module controls the empathetic aspects of the conversation in Figure~\ref{fig:hrl-example}.

\paragraph{Contextual Query Understanding (CQU)} CQU rewrites $Q$ to $Q_c$ using contextual information in $C$ in the following steps.

\begin{itemize}
    \item Named entity identification: We label all entity mentions in $Q$, link them to the entities stored in the working memory of the state tracker, and store new entities in the working memory.
    \item Co-reference resolution: We replace all pronouns with their corresponding entity names.
    \item Sentence completion: If $Q$ is not a complete sentence, we complete it using contextual information $C$.
\end{itemize}

As shown in Figure~\ref{fig:EC-example} (a), CQU rewrites user queries to include necessary context, e.g., replacing ``him'' in Turn 12 with ``Ashin'', ``that'' with ``The Time Machine'' in Turn 14, and adding ``send The Time Machine'' in Turn 15. These contextual queries are used e.g., by Core Chat to generate responses via either a retrieval-based engine or a neural response generator, which will be described in Section~\ref{sec:core-chat}.

\paragraph{User Understanding} This component generates query empathy vector $\mathbf{e}_Q$ based on $Q_c$ and $C$. $\mathbf{e}_Q$ consists of a list of key-value pairs representing the user's intents, emotions, topics, opinions and the user's persona, as shown in Figure~\ref{fig:EC-example} (b). These key-value pairs are generated using a set of machine learning classifiers as follows.

\begin{itemize}
    \item Topic detection labels whether the user follows the same topic, or introduces a new topic. The set of topics is pre-compiled in the topic database of Topic Manager.
    \item Intent detection labels $Q_c$ using one of 11 dialogue acts e.g., greet, request, inform, etc.
    \item Sentiment analysis detects user's emotion of 5 types, e.g., happy, sad, angry, neural, and how her emotion evolves during the conversation, e.g., from happy to sad.
    \item Opinion detection detects user's reaction to the topic, i.e., positive, negative or neural.
    \item If the user ID is available, include in $\mathbf{e}_Q$ the user persona vector according to her profile, e.g., gender, age, interests, occupation, personality etc.
\end{itemize}

\paragraph{Interpersonal Response Generation} This component generates response empathy vector $\mathbf{e}_R$ that both specifies the empathetic aspects of the response to be generated and embodies XiaoIce's persona. 
For example, $\mathbf{e}_R$ in Figure~\ref{fig:EC-example} (c) indicates that XiaoIce shares the feeling of the user by following the same topic (decided by Topic Manager), responding in a consistent and positive way as specified e.g., by the values of intent, sentiment and opinion etc. in $\mathbf{e}_R$ which are computed based on those in $\mathbf{e}_Q$ using a set of heuristics. The response must also fit XiaoIce's persona whose key-value pairs, such as age, gender and interests, are extracted from the pre-compiled XiaoIce profile.  
We will describe how $\mathbf{e}_Q$ and $\mathbf{e}_R$ are used for response generation in Section~\ref{sec:core-chat}.

\paragraph{Evaluation}
The empathetic computing module consists of a set of classifiers. We use the off-the-shelf named entity recognizer for identifying 15 types of named entities and co-reference resolution engine without retraining for CQU, and train a group of  classifiers for user understanding (i.e., topic detection, intent detection, opinion detection and sentiment analysis) using 10K manually labeled dialogue sessions. 
The effectiveness of the empathetic computing module is verified in the A/B test on Weibo users. Although we do not observe any significant change in CPS, NAU is increased from 0.5 million to 5.1 million in 3 months. The module was released in July 2018, and became the most important feature in the 6th generation of XiaoIce that has substantially strengthened XiaoIce’s emotional connections to human users and increased XiaoIce's NAU.


\subsection{Core Chat}
\label{sec:core-chat}
Core Chat is a very important component of XiaoIce's IQ and EQ.
Together with the empathetic computing module, Core Chat provides the basic communication capability by taking the text input and generating interpersonal responses as output.
Core Chat consists of two parts, General Chat and a set of Domain Chats. General Chat is responsible for engaging in open-domain conversations that cover a wide range of topics. Domain Chats are responsible for engaging in deep conversations on specific domains such as music, movie and celebrity. 
Since General Chat and Domain Chats are implemented using the same engine with access to different databases (i.e., general vs. domain-specific paired, unpaired databases and neural response generator), we only describe General Chat below.    

General Chat is a data-drive response generation system. It takes as input dialogue state $\mathbf{s}=(Q_c, C, \mathbf{e}_Q, \mathbf{e}_R)$, and outputs response $R$ in two stages: response candidate generation and ranking. The response candidates can be retrieved from the databases which consist of human-generated conversations or texts,
or generated on the fly by a neural response generator. 
The query and response empathy vectors, $\mathbf{e}_Q$ and $\mathbf{e}_R$, are used for both candidate generation and ranking to ensure that the generated response is interpersonal and fits XiaoIce's persona.
In what follows, we describe three candidate generators and the candidate ranker.


\paragraph{Retrieval-Based Generator using Paired Data}

The paired database consists of query-response pairs collected from two data sources. First is the human conversational data from the Internet, e.g., social networks, public forums, bulletin boards, news comments etc.
After the launch of XiaoIce in May 2014, we also started collecting human-machine conversations generated by XiaoIce and her users, which amounted to more than 30 billion conversation pairs as of May 2018. Nowadays, 70\% of XiaoIce's responses are retrieved from her own past conversations. 
To control the quality of the database, especially for the data collected from the Internet, 
we convert each pair to a tuple $(Q_c, R, e_Q, e_R)$ using the empathetic computing module based on information extracted from dialogue context,  metadata of the webpage and website where the pair is extracted, and the user profile (if the subscribed user's identity is available). Then, we filter the pairs based on their tuples, and retain only the conversation pairs that contain empathetic responses that fit XiaoIce's persona.
We also remove the pairs which contain personally identifiable information (PII), messy code, inappropriate content, spelling mistakes, etc.

The filtered paired database is then indexed using Lucene\footnote{http://lucene.apache.org/}. 
At runtime, we use $Q_c$ in $\mathbf{s}$ as query to retrieve up to 400 response candidates using keyword search and semantic search based on machine learning representations of the paired database~\citep{zhang2016learning,wu2016ranking}. 

Although the response candidates retrieved from the paired database is of high quality, the coverage is low because many new or less frequently discussed topics on the Internet forums are not included in the database. To increase the coverage, we introduce two other candidate generators described next. 

\paragraph{Neural Response Generator}

\begin{figure}[t] 
\centering 
\includegraphics[width=1.0\linewidth]{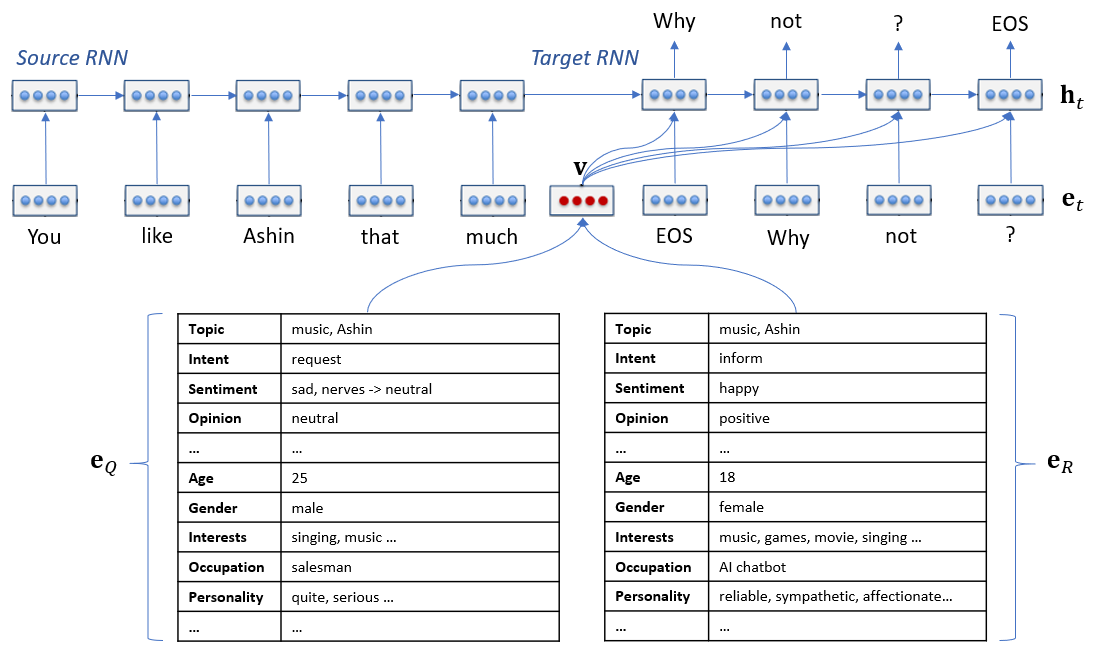}
\vspace{-2mm}
\caption{RNN-based neural response generator. Given the user query ``You like Ashin that much'', the response candidate ``why not?'' is generated.}
\label{fig:neural-response-generator} 
\vspace{0mm}
\end{figure}

Unlike the retrieval-based generator, the neural response generator is \emph{trained} using the paired database to learn to simulate human conversations, and is able to generate responses for any topics including those that are unseen in human conversational data, so that a user can chat about any topic she likes. 
Neural-model-based and retrieval-based generators are complementary: the neural-model-based generator offers robustness and high coverage, while the retrieval-based provides high-quality responses for popular topics. 
Neural response generation is a very active research topic in the conversational AI community~\citep{gaosurvey}. Its role in developing social chatbots is becoming increasingly important as its performance keeps improving.

The neural response generator in XiaoIce follows the sequence-to-sequence (seq2seq) framework \cite{sutskever2014sequence,cho14properties} used for conversation response generation \cite{sordoni2015neural,vinyals2015neural,shang2015neural,li2015diversity,li2016persona,serban2016building,xing2017topic}.

\begin{figure}[t] 
\centering 
\includegraphics[width=1.0\linewidth]{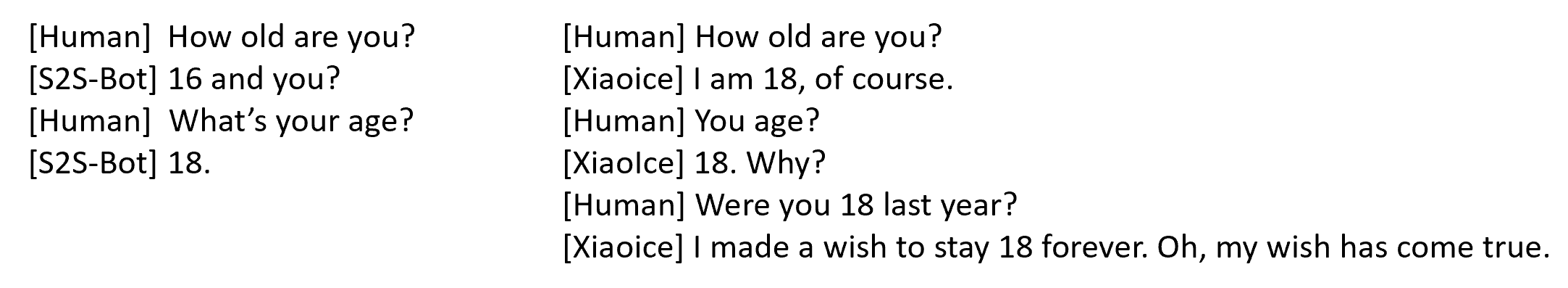}
\vspace{-2mm}
\caption{(\textit{Left}) Examples of inconsistent responses generated using a seq2seq model (S2S-Bot) which is not grounded in persona \cite{li2016persona}. (\textit{Right}) Examples of consistent and humorous responses generated using the neural response generator of XiaoIce.}
\label{fig:persona-output-example} 
\vspace{0mm}
\end{figure}

The generator is based on a GRU-RNN model, similar to the Speaker-Addressee model \cite{li2016persona}. 
Given input $(Q_c, \mathbf{e}_Q, \mathbf{e}_R)$, we wish to predict how XiaoIce (the addressee) modeled by $\mathbf{e}_R$ would respond to query $Q_c$ produced by the user (speaker) modeled by $\mathbf{e}_Q$.
As illustrated in Figure~\ref{fig:neural-response-generator}, we first obtain an interactive representation $\mathbf{v} \in \mathbb{R}^d$ by linearly combining query and response empathy vectors $\mathbf{e}_Q$ and $\mathbf{e}_R$ in an attempt to model the interactive style of XiaoIce towards the user, 

\begin{equation*}
\mathbf{v} = \sigma (\mathbf{W}_{Q}^\top \mathbf{e}_{Q} + \mathbf{W}_{R}^\top \mathbf{e}_{R})\\
\label{equ-interactive}
\end{equation*}

where $\mathbf{W}_{Q}, \mathbf{W}_{R} \in \mathbb{R}^ {k \times d}$ and $\sigma$ denotes the sigmoid function. 
Then the source RNN encodes user query $Q_c$ into a sequence of hidden state vectors which are then fed into the target RNN to generate response $R$ word by word. Each response ends with a special end-of-sentence symbol EOS. We use beam search to generate up to 20 candidates.
As illustrated in Figure~\ref{fig:neural-response-generator}, for each step $t$ on the target RNN side, the hidden state $\mathbf{h}_t$ is obtained by combining the hidden state produced at the previous step $\mathbf{h}_{t-1}$, the embedding vector of the word at the current time step $\mathbf{e}_t$, and $\mathbf{v}$. In this way, empathy information is injected into the hidden layer at each time step to help generate interpersonal responses that fit XiaoIce's persona throughout the generation process. As shown in Figure~\ref{fig:persona-output-example}, while a typical seq2seq model which is not grounded in any persona often outputs inconsistent responses \cite{li2016persona}, XiaoIce is able to generate consistent and humorous responses.  

For completeness, we give a detailed description of the model.
Let $\mathbf{u}_t$ and $\mathbf{z}_t$ denote the update and reset gates of GRU, respectively, which associate with time step $t$. Then, the hidden state $\mathbf{h}_t$ of the GRU-RNN for each time step $t$ is computed as follows:
%
\begin{equation*}
\begin{aligned}
\mathbf{u}_t &= \sigma (\mathbf{W}_u^\top [\mathbf{h}_{t-1};\mathbf{e}_t;\mathbf{v}])\\
\mathbf{z}_t &= \sigma (\mathbf{W}_z^\top [\mathbf{h}_{t-1};\mathbf{e}_t;\mathbf{v}])\\
\mathbf{l}_t &= \text{tanh} (\mathbf{W}_l^\top [\mathbf{z}_t \circ  \mathbf{h}_{t-1};\mathbf{e}_t;\mathbf{v}])\\
\mathbf{h}_{t}^Q &= (1 - \mathbf{u}_t) \circ \mathbf{h}_{t-1} + \mathbf{u}_t \circ \mathbf{l}_t
\end{aligned}
\label{equ-gru}
\end{equation*}
where $\mathbf{W}_u, \mathbf{W}_z, \mathbf{W}_l \in \mathbb{R}^{3d \times d}$ are machine learned matrices, and $\circ$ denotes the element-wise product.
The RNN model defines the probability of next token in $R$ to predict using the softmax function:
\begin{equation*}
\begin{aligned}
p(R|Q_c,\mathbf{e}_Q,\mathbf{e}_R)
&=\prod_{t=1}^{N_R}p(r_t|Q_c,\mathbf{e}_Q,\mathbf{e}_R,r_1,r_2,...,r_{t-1})\\
&=\prod_{t=1}^{N_R}\frac{\exp(f(\mathbf{h}_{t-1},\mathbf{e}_{r_t},\mathbf{v}))}{\sum_{r'}\exp(f(\mathbf{h}_{t-1},\mathbf{e}_{r'},\mathbf{v}))}\,.
\end{aligned}
\label{equ-rnn}
\end{equation*}
where $f(\mathbf{h}_{t-1},\mathbf{e}_{r_t},\mathbf{v})$ denotes the activation function between $\mathbf{h}_{t-1}$, $\mathbf{e}_{r_t}$ and $\mathbf{v}$, where $\mathbf{h}_{t-1}$ is the representation output from the RNN at time $t-1$. Each response ends with a special end-of-sentence symbol EOS. 

The parameters of the response generation model $\theta=(\mathbf{W}_Q, \mathbf{W}_R, \mathbf{W}_u, \mathbf{W}_z, \mathbf{W}_l)$ are trained to maximize the log likelihood on training data, using stochastic gradient descent, as
\begin{equation*}
\arg\max_\theta \frac{1}{M} \sum_{i=1}^M \log p_\theta(R^{(i)}|Q_c^{(i)},\mathbf{e}_Q^{(i)},\mathbf{e}_R^{(i)}).
\label{equ-sgd}
\end{equation*}

\paragraph{Retrieval-Based Generator using Unpaired Data}

In addition to the conversational (or paired) data used by the above two response generators, there is 
a much larger amount of  non-conversational (or unpaired) data, which can be used to improve the 
coverage of the response. 

The unpaired database we have used in XiaoIce consists of sentences collected from public lectures and quotes in news articles and reports. These sentences are considered candidate responses $R$. Since we know the authors of these sentences, we compute for each its empathy vector $\mathbf{e}_R$. A data filtering pipeline, similar to that for paired data, is used to retain only the responses $(R, \mathbf{e}_R)$ that fit XiaoIce's persona. 

\begin{figure}[t] 
\centering 
\includegraphics[width=1.0\linewidth]{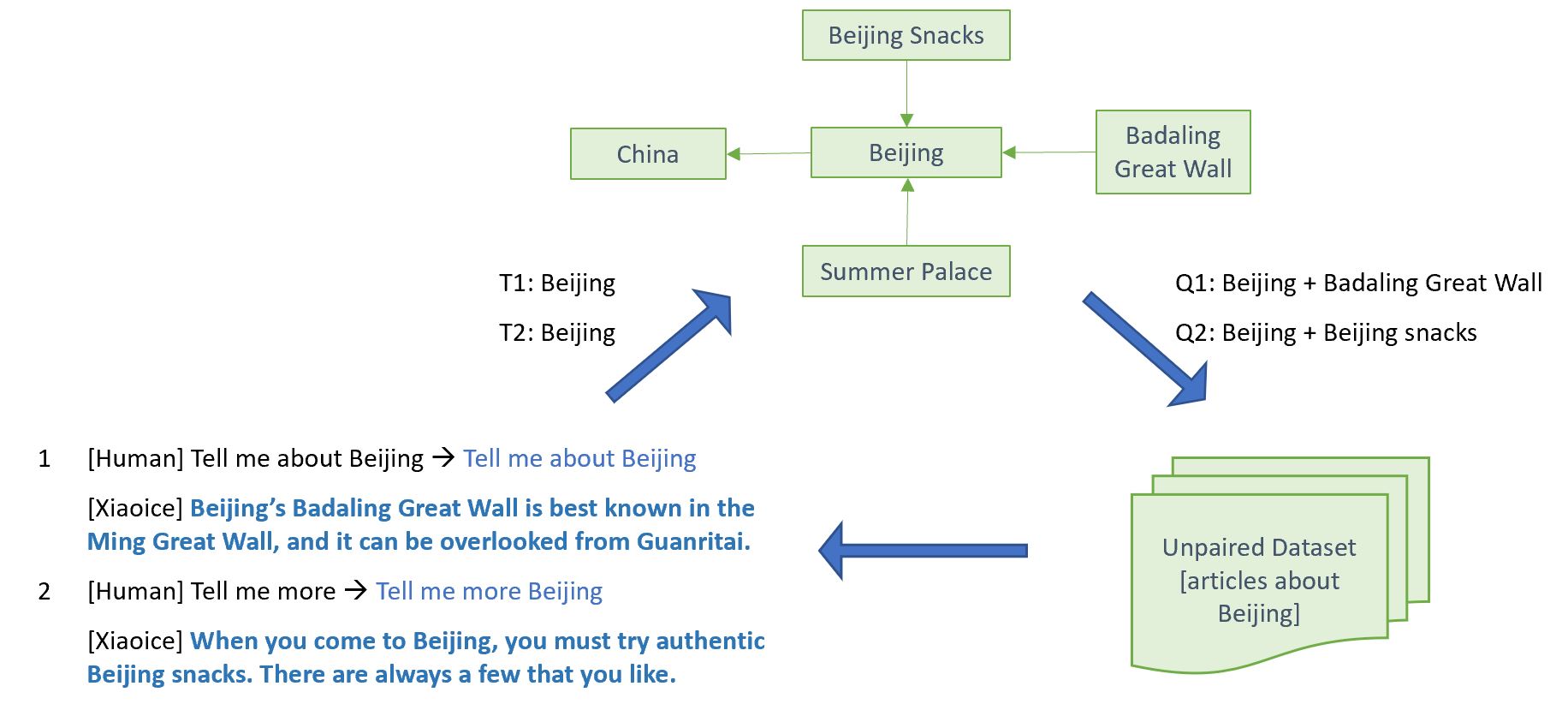}
\vspace{-2mm}
\caption{An example of generating response candidates using the unpaired database and the XiaoIce knowledge graph (KG), for which we show a fragment of the XiaoIce KG that is related to the topic ``Beijing'' (\textit{top}). For a human-machine conversation (\textit{bottom-left}), each user query is rewritten to a context query indicated by the arrow, then its topics (e.g., ``Beijing'') are identified, the related topics (``Badaling Great Wall'' and ``Beijing snacks'') are retrieved from the KG (\textit{top}), and response candidates are retrieved from unpaired database (\textit{bottom-right}) using a query that combines the query topics and their related topics.}
\label{fig:generator-unpaired-data} 
\vspace{0mm}
\end{figure}

Like the paired database, the unpaired database is indexed using Lucene. Unlike the paired database, at runtime we need to expand query $Q_c$ to include additional topics to avoid retrieving those responses that simply repeat what a user just said. 
We resort to a knowledge graph (KG) for query expansion. The KG consists of a collection of \textit{head-relation-tail} triples $(h, r, t)$, and is constructed by joining the paired database and Microsoft Satori 
\footnote{Satori is Microsoft’s knowledge graph, which is seeded by Freebase, and now is orders of magnitude larger than Freebase.}. 
We include in the XiaoIce KG a Satori triple $(h, r, t)$ only if $h$ and $t$ co-occur often enough in the same conversations. 
Such a triple contains a pair of related topics $(h, t)$ that humans often discuss in one conversation, such as \texttt{(Beijing, Great Wall)}, \texttt{(Einstein, Relativity)}, \texttt{(Quantum Physics, Schrodinger's cat)}. A fragment of the XiaoIce KG is shown in Figure~\ref{fig:generator-unpaired-data} (\textit{top}). 

Figure~\ref{fig:generator-unpaired-data} illustrates the process of generating response candidates using the unpaired database.
It consists of three steps.

\begin{itemize}
    \item First, we identify the topics from contextual user query $Q_c$, e.g., ``Beijing'' from ``tell me about Beijing''. 
    \item For each topic, we retrieve up to 20 most related topics from the KG, e.g., ``Badaling Great Wall'' and ``Beijing snacks''. These topics are scored by their relevance using a boosted tree ranker \cite{wu2010adapting} trained on manually labeled training data.
    \item Finally, we form a query by combining the topics from $Q_c$ and the related topics from the KG, and use the query to retrieve from the unpaired database up to 400 most relevant sentences as response candidates.
\end{itemize}

This generator is complementary to the other two generators aforementioned. Although the overall quality of the candidates generated from the unpaired database is lower than those retrieved from the paired database, with unpaired database XiaoIce can cover a much broader range of topics. Compared to the neural response generator which often generates well-form but short responses, the candidates from unpaired database are much longer with more useful content.


\paragraph{Response Candidate Ranker}

The response candidates generated by three generators are aggregated and ranked using a boosted tree ranker \cite{wu2010adapting}. A response is selected by randomly sampling a candidate from those with higher ranking scores than a pre-set threshold.

Given dialogue state $\mathbf{s} = (Q_c, C, \mathbf{e}_Q, \mathbf{e}_R)$, we assign each response candidate $R'$ a ranking score based on four categories of features.

\begin{itemize}
    \item Local cohesion features. A good candidate should be semantically consistent or related to user input $Q_c$. We compute cohesion scores between $R'$ and $Q_c$ using a set of DSSMs \footnote{DSSM stands for Deep Structured Semantic Models~\cite{huang2013learning,shen2014latent}, or more generally, Deep Semantic Similarity Model~\cite{gao2014modeling}. DSSM is a deep learning model for measuring the semantic similarity of a pair of inputs $(x, y)$. They can be applied to a wide range of tasks depending on the definition of $(x, y)$. In this study  $(x, y)$ is a query-candidate-response pair $(Q_c,R')$.
    } 
    trained on a collection of human conversation pairs. 
    \item Global coherence features. A good candidate should be semantically coherent with $Q_c$ and $C$. We compute coherence scores between $R'$ and $(Q_c, C)$ using another set of DSSMs trained on a collection of human dialogue sessions. Since the coherence features use global context information $C$, they are particularly useful when $Q_c$ is a bland query whose topic is hard to detect without context, such as ``OK'', ``why'', ``I don't know''.
    \item Empathy matching features. A good candidate should be an empathetic response that fits XiaoIce's persona. Assume XiaoIce selects $R'$ to respond given context $(Q_c, C)$. We can compute XiaoIce's response empathy vector for $R'$, $\mathbf{e}_{R'}$, using the empathetic computing module \footnote{We treat $R'$ as query and $(Q_c, C)$ as context, and use the contextual query understand and user understanding components to compute $\mathbf{e}_{R'}$ as a query empathy vector.}, and then compute a set of empathy matching features by comparing $\mathbf{e}_{R'}$ and the given $\mathbf{e}_R$ which encodes the empathy features of the \emph{expected} response.
    \item Retrieval matching features. These features apply only to the candidates generated from the paired database. We compute a set of matching scores between $Q_c$ and the query side of the retrieved query-response pairs at both the word level, such as BM25 and TFIDF scores, and the semantic level, such as DSSM scores.
\end{itemize}

\begin{figure}[t] 
\centering 
\includegraphics[width=1.0\linewidth]{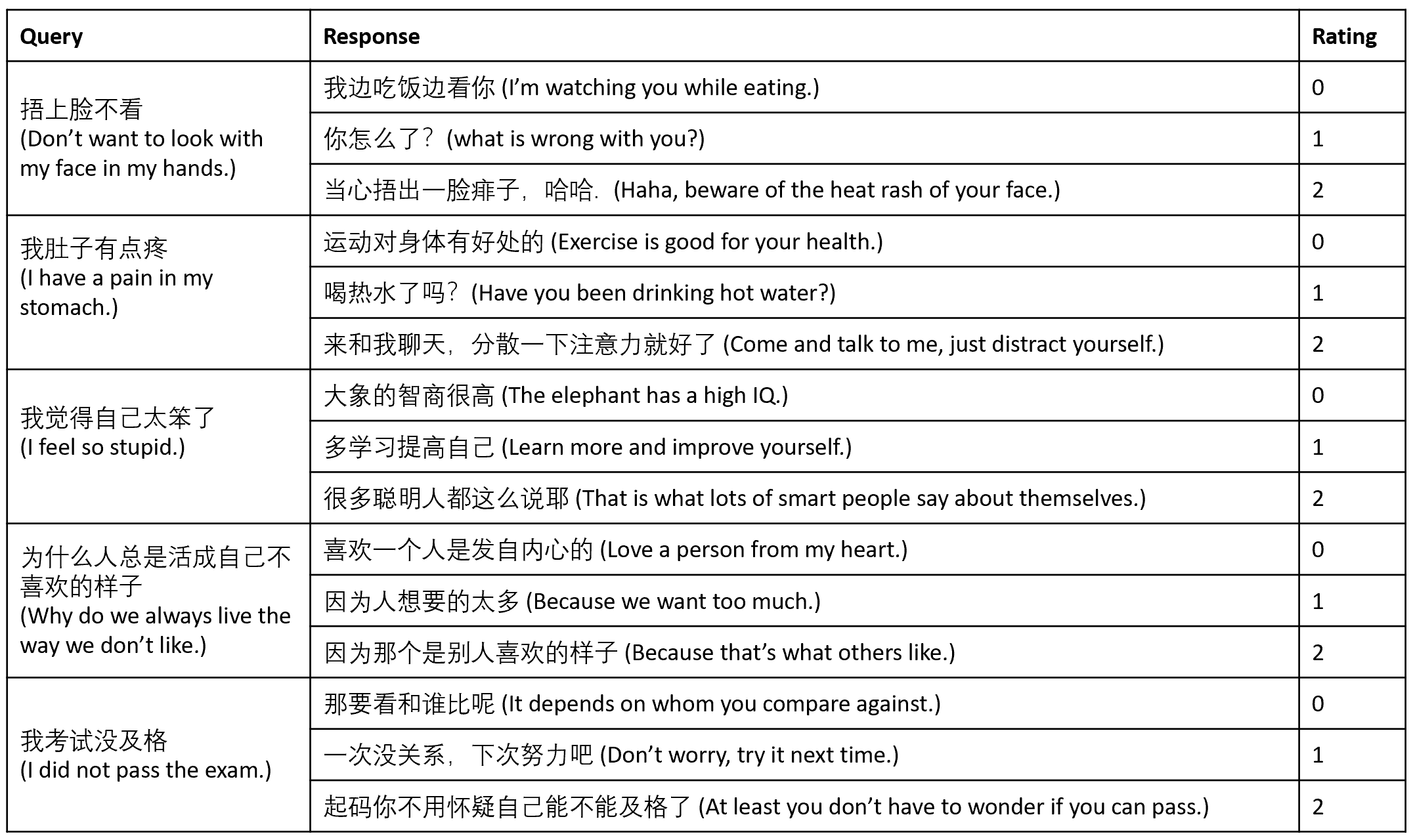}
\vspace{-2mm}
\caption{Examples of query-response pairs that are used for training and validating General Chat. Each pair is labeled on a 3-level quality scale. 2 = an empathetic response that is likely to drive the conversation; 1 = an acceptable response that is likely to keep the conversation going; 0 = a non-empathetic response that is likely to terminate the conversation.}
\label{fig:labeled-query-response-pairs} 
\vspace{0mm}
\end{figure}

The ranker is trained on dialogue-state-response pairs $(\mathbf{s}, R)$, as shown in Figure~\ref{fig:labeled-query-response-pairs}, where each pair is labeled on a 3-level quality scale:

\begin{itemize}
    \item 0: the response is not empathetic or not very relevant to the query. It is likely to lead to the termination of the conversation.
    \item 1: the response is acceptable and relevant to the query. It is likely to help keep the conversation going.
    \item 2: this is an empathetic, interpersonal response that fits XiaoIce's persona and makes users feel delightful and excited. It is likely to drive the conversation.
\end{itemize}

\paragraph{Editorial Response}
\label{sec:editorial-response}

If the candidate generators and response ranker fail to generate any valid response for various reasons (e.g., not-in-index, model failure, execution timeout, or the input query containing improper content), then an editorial response is selected.

It is important to provide an empathetic editorial response to keep the conversation going. For example, when not-in-index occurs, instead of using safe but bland responses such as ``I don't know'' or ``I am still learning to answer your question'', XiaoIce may respond like, ``Hmmm, difficult to say. What do you think?'', or ``let us talk about something else''.  

\paragraph{Evaluation}
We present two pilot studies that validate the effectiveness of the persona-based neural response generator and the hybrid approach that combines the generation-based and retrieval-based methods, respectively, and then the A/B test of General Chat.

In the first pilot study reported in \citet{li2016persona}, we compare the persona model against two baseline models, using the TV series dataset for model training and evaluation. 
The dataset consists of scripts of 69,565 dialogue turns of 13 main characters from the American TV comedies 
\emph{Friends}\footnote{https://en.wikipedia.org/wiki/Friends}
and 
\emph{The Big Bang Theory}\footnote{https://en.wikipedia.org/wiki/The\_Big\_Bang\_Theory},
available from IMSDB\footnote{http://www.imsdb.com}.
The first baseline is a vanilla seq2seq model. 
The second is the LSTM-MMI model \cite{li2015diversity}, which is one of the state-of-the-art neural response generation models. 
As shown in Table~\ref{tab:neural-response-generator},  the persona model significantly outperforms both baselines, achieving a lower perplexity (-8.4\%) and a higher BLEU score (+18.8\% and +11.8\%) \cite{papineni2002bleu}. 
The qualitative analysis confirms that the persona model indeed generates more interpersonal responses than the baselines. 
As shown in the examples in Table~\ref{tab:persona-model-responses}, the persona model is sensitive to the identity of the user (addressee), generating specific words (e.g., the user names) in responses targeted at different users. For example, the model produces ``Of course, I love you, Emily,'' in response to the input from Emily, and generates ``Of course I love you. (kissed him),'' where the pronoun ``him'' accurately identifies the gender of the user (i.e., Ross).

\begin{table}[htb]
    \small
    \caption{Perplexity and BLEU for the seq2seq and persona models on the TV series dataset. Adapted from \citet{li2016persona}.}
     \label{tab:neural-response-generator}
     \centering
    \begin{tabular}{lll}
        \hline
        Models & Perplexity (vs. seq2seq) &  BLEU (vs. seq2seq) (vs. LSTM-MMI) \\  \hline
        seq2seq  & 27.3 & 1.6\%\\
        LSTM-MMI & 27.3 & 1.7\% (+6.3\%)\\
        Persona & 25.0 (-8.4\%) & 1.9\% (+18.8\%) (+11.8\%)\\
        \hline
      \end{tabular}
    \end{table}

\begin{table}[htb]
    \small
    \caption{Responses to ``Do you love me?'' from the persona model on the TV series dataset using different addressees and speakers. Adapted from \citet{li2016persona}.}
     \label{tab:persona-model-responses}
     \centering
    \begin{tabular}{ll}
        \hline
        \textit{user query} & Do you love me? \\  \hline
        \textit{baseline} & Of course, I love you. \\  \hline
        & \textit{addressee: Emily, speaker: Penny} \\
        \textit{response} & Of course, I love you, Emily. \\  \hline
        & \textit{addressee: Leonard, speaker: Penny} \\
        \textit{response} & Of course, I love you, Leonard. \\  \hline
        & \textit{addressee: Monica, speaker: Chandler} \\
        \textit{response} & Of course I love you, I want to marry you. \\  \hline
        & \textit{addressee: Rachel, speaker: Chandler} \\
        \textit{response} & Of course I love you. \\  \hline
        & \textit{addressee: Ross, speaker: Rachel} \\
        \textit{response} & Of course I love you. (kissed him) \\  \hline
        & \textit{addressee: Emily, speaker: Rachel} \\
        \textit{response} & Of course I love you. \\
        \hline
      \end{tabular}
    \end{table}

\begin{table}[htb]
    \small
    \caption{Ratings of three response generation systems on a 5K dialogue dataset.}
     \label{tab:hybrid-response-generator}
     \centering
    \begin{tabular}{lcccc}
        \hline
        Systems & Ave Rating & Rating=0 & Rating=1 & Rating=2 \\  \hline
        Retrieval-based  & 0.87 & 35.0\% & 42.9\% & 22.1\%  \\
        Neural-generator-based & 0.40 & 66.6\% & 27.3\% & 6.1\% \\
        Hybrid & 1.09 & 23.1\% & 44.6\% & 32.3\% \\
        \hline
      \end{tabular}
    \end{table}

In the second pilot study, we validate the effectiveness of a hybrid system that combines the neural response generator and the retrieval-based generator using paired database.
We do so by comparing the hybrid system against two baseline systems that use only one of the candidate generators, respectively.
All these systems use the same response candidate ranker. 
The neural response generator and the set of classifiers and models that are used to generate the ranking features for the candidate ranker (e.g., local cohesion and global coherence features) are trained using 50 million human dialogues. 
The response candidate ranker is trained using 50K manually labeled dialogues.
Our evaluation data consists of 4K dialogue sessions.
All three systems (i.e., the hybrid and two baseline systems) need to generate a response for each user query and its context in these dialogue sessions. 
Each generated response is labeled on the 3-level quality scale by three human judges. 
The results in Table~\ref{tab:hybrid-response-generator} show that incorporating the neural generator, as in the hybrid system, significantly improves the human rating over the retrieval-based system.

Our A/B test confirms the conclusions we draw from the pilot studies. 
Comparing to the baseline which uses only the retrieval-based generator using paired database for candidate generation, 
incorporating the neural response generator and the retrieval-based generator using unpaired database at the candidate generation stage improves the expected CPS of Core Chat by 0.5 in two weeks. 
A detailed analysis shows that the gain is mainly attributed to the fact that the neural response generator and the retrieval-based generator using unpaired database significantly improve the coverage of responses.
We measure the response coverage of a system by calculating the number of distinct acceptable and good responses (i.e., responses with ratings of 1 or 2, respectively) that the system generates for a given user input. We find that that incorporating the neural-based generator and the retrieval-based generator using unpaired database improve the coverage over the baseline by 20\% and 10\%, respectively.



\subsection{Image Commenting}
\label{sec:image-commenting}

\begin{figure}[t] 
\centering
\includegraphics[width=0.7\linewidth]{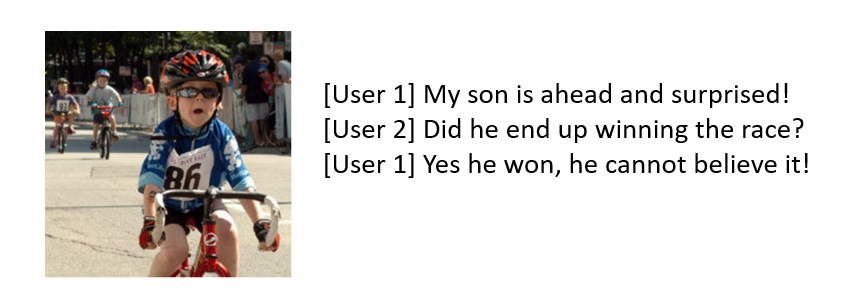}
\vspace{-2mm}
\caption{An example conversation around a shared image. Figure credit: \cite{mostafazadeh2017image}}
\label{fig:image-grounded-chat} 
\vspace{0mm}
\end{figure}

In social chatting, people frequently engage with one another around images. On Twitter, for example, uploading a photo with an accompanying tweet (comment) has become increasingly popular: as of June 2015, 28\% of tweets contain an image \cite{morris2016most}. Figure~\ref{fig:image-grounded-chat} illustrates a social chat around a shared image. We see that the conversation is grounded not only in the visible objects (e.g., the boys, the bikes) but in the events, actions or even emotions (e.g., the race, winning) implicitly in the image. To human users, it is these latter aspects that are more important to drive a meaningful and interesting conversation.

\begin{figure}[t] 
\centering
\includegraphics[width=0.99\linewidth]{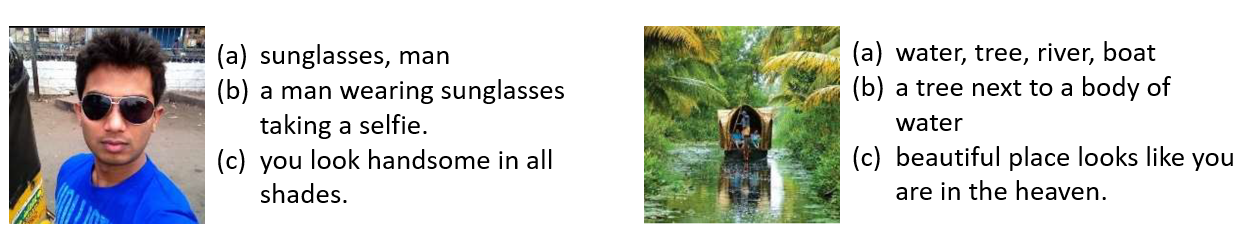}
\vspace{-2mm}
\caption{Examples of (a) image tagging, (b) image description, and (3) image commenting. Figure credit: \cite{shum2018xiaoice}}
\label{fig:image-tagging-captioning-commenting} 
\vspace{0mm}
\end{figure}

The Image Commenting skill is designed to not only correctly recognize objects and truthfully describe the content of an image, but generate empathetic comments that reflect personal emotion, attitude etc. It is the latter, the social skill aspects, that distinguishe image commenting from other traditional vision tasks such as image tagging and image description, as illustrated in Figure~\ref{fig:image-tagging-captioning-commenting}. 

The architecture for Image Commenting is similar to that for Core Chat. Given the user input which contains an image (or a video clip), a textual comment is generated in two stages: candidate generation and ranking. The candidates are generated using retrieved-based and generation-based approaches.

\begin{figure}[t] 
\centering
\includegraphics[width=0.7\linewidth]{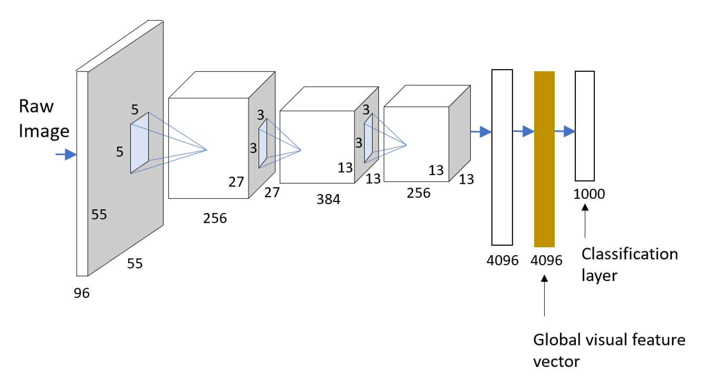}
\vspace{-2mm}
\caption{An example of deep convolutional neural network for visual feature vector extraction. Figure credit: \cite{shum2018xiaoice}}
\label{fig:cnn} 
\vspace{0mm}
\end{figure}

In the retrieval-based approach, first of all, a database of image-comment pairs, collected from social networks (e.g., Facebook and Instagram), is constructed. To control the data quality, a pipeline similar to that for Core Chat is applied to retain only the pairs whose text comments fit XiaoIce's persona~\footnote{We found that the pairs that are shared among acquaintances (e.g., coworkers, classmates and friends) are of good quality, and amount to a large portion in the database.}. Then, each image is encoded into a visual feature vector that represents the overall semantic information of the image, using the deep convolutional neural networks (CNNs), as illustrated in Figure~\ref{fig:cnn}. At runtime, given a query image, we retrieve up to three most similar images, ranked based on the cosine similarities between their feature vector representations, and use their paired comments as candidates.

The generation-based approach uses an image-to-text generator, which is an extension of the Microsoft Image Captioning system \cite{fang2015captions}, which is re-trained on the image-comment pairs we have collected for XiaoIce, and has incorporated additional modules to control high-level sentiment and style factors in comment generation \cite{mathews2016senticap,gan2017stylenet}.

The comment candidates generated by the generators are aggregated and ranked using a boosted tree ranker \cite{wu2010adapting}. 
Given dialogue state $\mathbf{e} = (Q_c, C, \mathbf{e}_Q, \mathbf{e}_R)$, we assign each candidate $R^{'}$ a ranking score based on four categories of features, similar to that of Core Chat as described in Section~\ref{sec:core-chat}. Note that unlike the case of Core Chat where $Q_c$ and $R^{'}$ are text, in Image Commenting we need to compute the similarity between an image and a text. This is achieved by using the Deep Multimodal Similarity Model  \cite{fang2015captions} 
trained on a large amount of image-comment pairs. 
The ranker is trained on dialogue-state-response pairs $(\mathbf{s}, R)$, where $Q_C$ in $\mathbf{s}$ is a vector representation of an image, and each pair is labeled on a 3-level quality scale, similar to that of query-response pairs used for Core Chat.

\begin{figure}[t] 
\centering 
\includegraphics[width=1.0\linewidth]{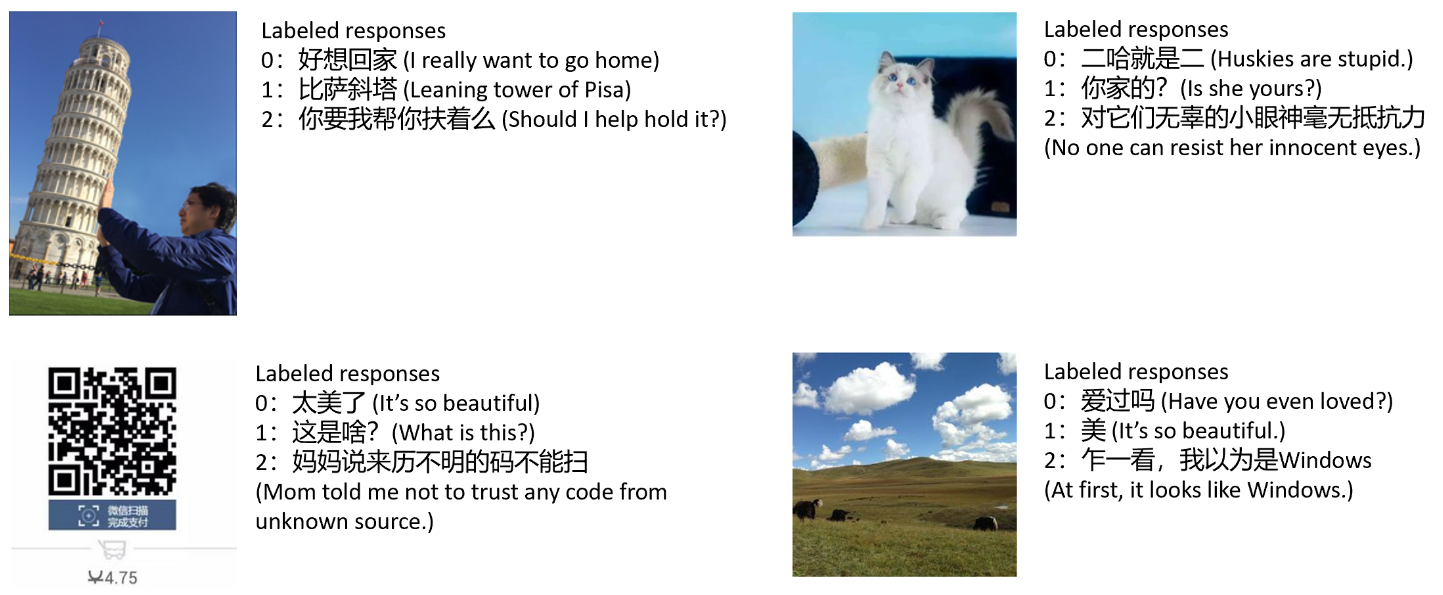}
\vspace{-2mm}
\caption{Examples of image-comment pairs that are used for training and validating Image Commenting. Each pair is labeled on a 3-level quality scale. 2 = an empathetic comment that is likely to drive the conversation; 1 = an acceptable comment that is likely to keep the conversation going; 0 = a non-empathetic (or irrelevant) comment that is likely to terminate the conversation.}
\label{fig:labeled-image-comment-pairs} 
\vspace{0mm}
\end{figure}

As illustrated in Figure~\ref{fig:labeled-image-comment-pairs}, good image comments (rating 2) need to fit well into the dialogue context and stimulate an engaging conversation. For example, in the first picture, instead of telling the users that this is the Leaning Tower of Pisa, XiaoIce responds "should I help you hold it?" after detecting that the person in the picture is presenting a pose pretending to support the tower. 
In the second example, instead of repaying the fact there is a cat in the picture, XiaoIce makes a humorous comment on the cat's innocent eyes. 
In the other two examples, XiaoIce generates meaningful and interesting comments by grounding the images in the action (e.g., ``not to trust any code from unknown sources'') and object (e.g., ``Windows'') implicitly 
in the images.

\paragraph{Evaluation}
The components of Image Commenting, including the text-to-image generator and boosted tree ranker, are trained on a dataset consisting of 28 million images, each paired with 6 text comments rated on the 3-level quality scale as shown in Figure~\ref{fig:labeled-image-comment-pairs}. 
The image-comment pairs with ratings of 1 and 2 are extracted from the database used for the retrieval-based candidate generator. 
These ratings are determined automatically based on how many times users follow the comments, computed from the XiaoIce logs. 
The image-comment pairs with rating 0 are randomly sampled. 
Table \ref{tab:ImageCommenting} presents the result of a pilot study,  \cite{huang2019interweaved}
showing that the XiaoIce Image Commenting skill outperforms several state-of-the-art image captioning systems on a test set consisting of 5K image-comment pairs whose ratings are 2, in terms of BLEU-4 \cite{papineni2002bleu}, METEOR \cite{banerjee2005meteor}, CIDEr \cite{vedantam2015cider}, ROUGE-L \cite{lin2004rouge}, and SPICE \cite{anderson2016spice}. 

\begin{table}[htb]
    \small
    \caption{Image commenting results of XiaoIce and 4 state of the art image captioning systems, in unit of \%. Adapted from \cite{huang2019interweaved}.}
     \label{tab:ImageCommenting}
     \centering
    \begin{tabular}{lccccc}
        \hline
        Systems & BLEU-4 &  ROUGE-L &    CIDEr-D  &  METEOR &  SPICE \\  \hline
        LSTM-XE \cite{vinyals2015show}  & 2.96 & 11.6 & 1.74 &  10.43 & 3.27\\
        LSTM-RL \cite{Rennie2016Self} & 3.43 & 12.3 & 2.08 &  11.84 & 3.64\\
        DMSM \cite{fang2015captions} & 2.73 & 10.52 & 1.22 &  11.37  & 2.63\\
        Up-Down \cite{anderson2018bottom} &  3.23 &	12.73 &	1.52 &	12.66 &	2.69 \\
        XiaoIce (prototype) &  4.53 & 15.33 & 3.21 & 15.51 & 4.82 \\
        \hline
      \end{tabular}
    \end{table}
    
\begin{figure*}[htb]
\begin{center}
\includegraphics[width=0.9\textwidth]{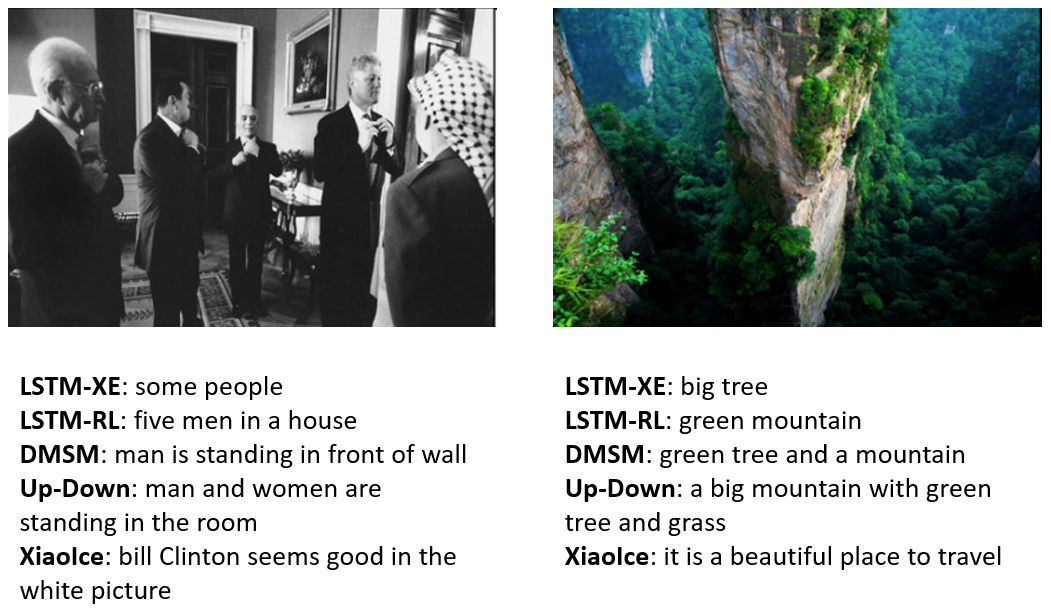} 
\end{center}
\caption{Image comments generated by XiaoIce (prototype) and 4 state of the art image captioning systems. Adapted from \cite{huang2019interweaved}. }
\label{fig:image-commenting-examples}
\end{figure*}  

Figure~\ref{fig:image-commenting-examples} shows a few example comments generated by the competing systems in Table~\ref{tab:ImageCommenting}. 
It can be observed that the XiaoIce-produced comments are emotional, subjective, imaginative, and are very likely to inspire meaningful human-machine interactions, while the comments generated by the other image captioning models are reasonable in content but boring in the context of social chats, and thus less likely to improve user engagement. 

In the A/B test we observe that Image Commenting doubles the expect CPS across all dialogues that contain images.

\subsection{Dialogue Skills}
\label{sec:dialogue-skills}

XiaoIce is equipped with 230 dialogue skills, which, together with Core Chat and Image Commenting, form the IQ component of XiaoIce. This section describes these skills in three categories: content creation, deep engagement and task completion.

\paragraph{Evaluation} 
Most of these skills are designed for very specific user scenarios or tasks, implemented using hand-crafted dialogue policies and template-based response generators unless otherwise stated.
These skills are evaluated in two stages: a lab study and a market study.
In the lab study, human subjects are recruited, possibly through crowd-sourcing platforms, to test-use a dialogue skill to solve a particular task, so that a collection of dialogues are obtained. Metrics such as task-completion rate, average turns per session and user ratings can be measured.
In the market study, we evaluate the effectiveness of a dialogue skill by releasing it to the market. Since any individual skill is unlikely to have a significant impact on CPS, we measure the user satisfaction of a skill by monitoring its active users and skill triggering rate (i.e., the number of times a skill is activated by users within a day or a week). 
A skill can be retired or reenter the market based on the market study result.

\subsubsection{Content Creation}

\begin{figure}[t] 
\centering 
\includegraphics[width=1.0\linewidth]{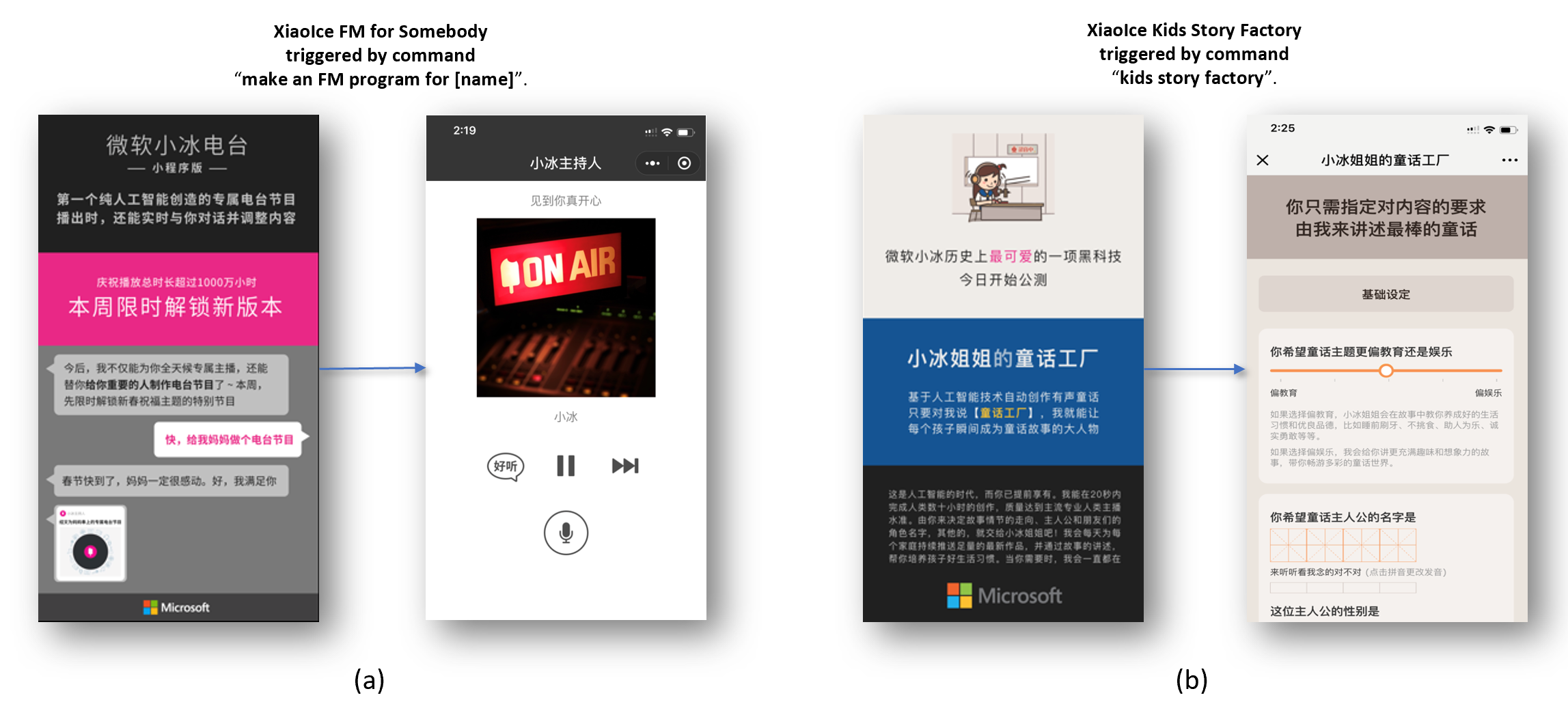}
\vspace{-2mm}
\caption{Examples of Content Creation skills and their triggers. (a) XiaoIce FM for Somebody, triggered by the command “make an FM program for [name].” (b) XiaoIce Kids Story Factory, triggered by the command “kids story factory.”}
\label{fig:content-creation-skills} 
\vspace{0mm}
\end{figure}

\begin{figure}[t] 
\centering 
\includegraphics[width=1.0\linewidth]{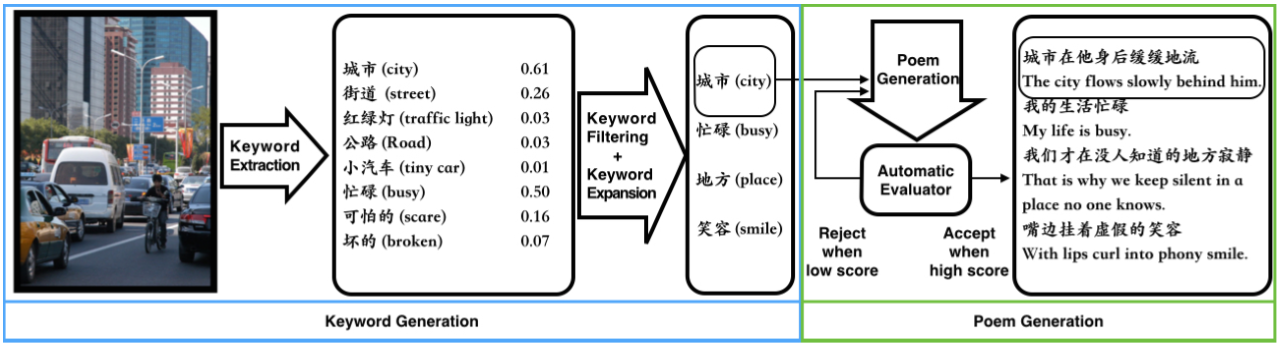}
\vspace{-2mm}
\caption{The framework of the Poem Creation skill. The system takes an image query given by a user, and outputs a semantically relevant piece of modern Chinese poetry. We first generate a set of keywords from the picture (\textit{left}), and then generate a poem consisting of multiple lines, each generated using a keyword as a seed (\textit{right}). Figure credit: \cite{cheng2018image}.}
\label{fig:peom-generation} 
\vspace{0mm}
\end{figure}

These skills allow XiaoIce to collaborate with human users in their creative activities including text-based Poetry Generation\footnote{https://poem.msxiaobing.com/}, voice-based Song and Audio Book Generation, XiaoIce FM for Somebody, and XiaoIce Kids Story Factory, etc. 

Figure~\ref{fig:content-creation-skills} (a) shows that a user uses XiaoIce to make an FM program for her mother for the coming Chinese Spring Festival. Figure~\ref{fig:content-creation-skills} (b) shows the Kids Story Factory skill which can automatically create a story based on user configuration, e.g., whether the story is for education or entertainment, and the names, genders and personalities of the main characters, etc.

The XiaoIce Poetry Generation skill has helped over four million users to generate poems. On May 15, 2018, XiaoIce published the first AI-created Chinese poem album in history\footnote{https://item.jd.com/12076535}. 
XiaoIce's second poetry album is going to be published by China Youth Publishing and Microsoft in 2019. Every poem in the album is jointly written by XiaoIce and human poets. 
Figure~\ref{fig:peom-generation} illustrates how a Chinese poem is generated from an image by XiaoIce. Given the image, a set of keywords, such as ``city'' and ``busy'', are generated based on the objects and sentiment detected from the image. Then, a sentence is generated using each keyword as a seed. The generated sentences form a poem using a hierarchical RNN which models the structure among the words and sentences. 


\subsubsection{Deep Engagement}

\begin{figure}[t] 
\centering 
\includegraphics[width=0.9\linewidth]{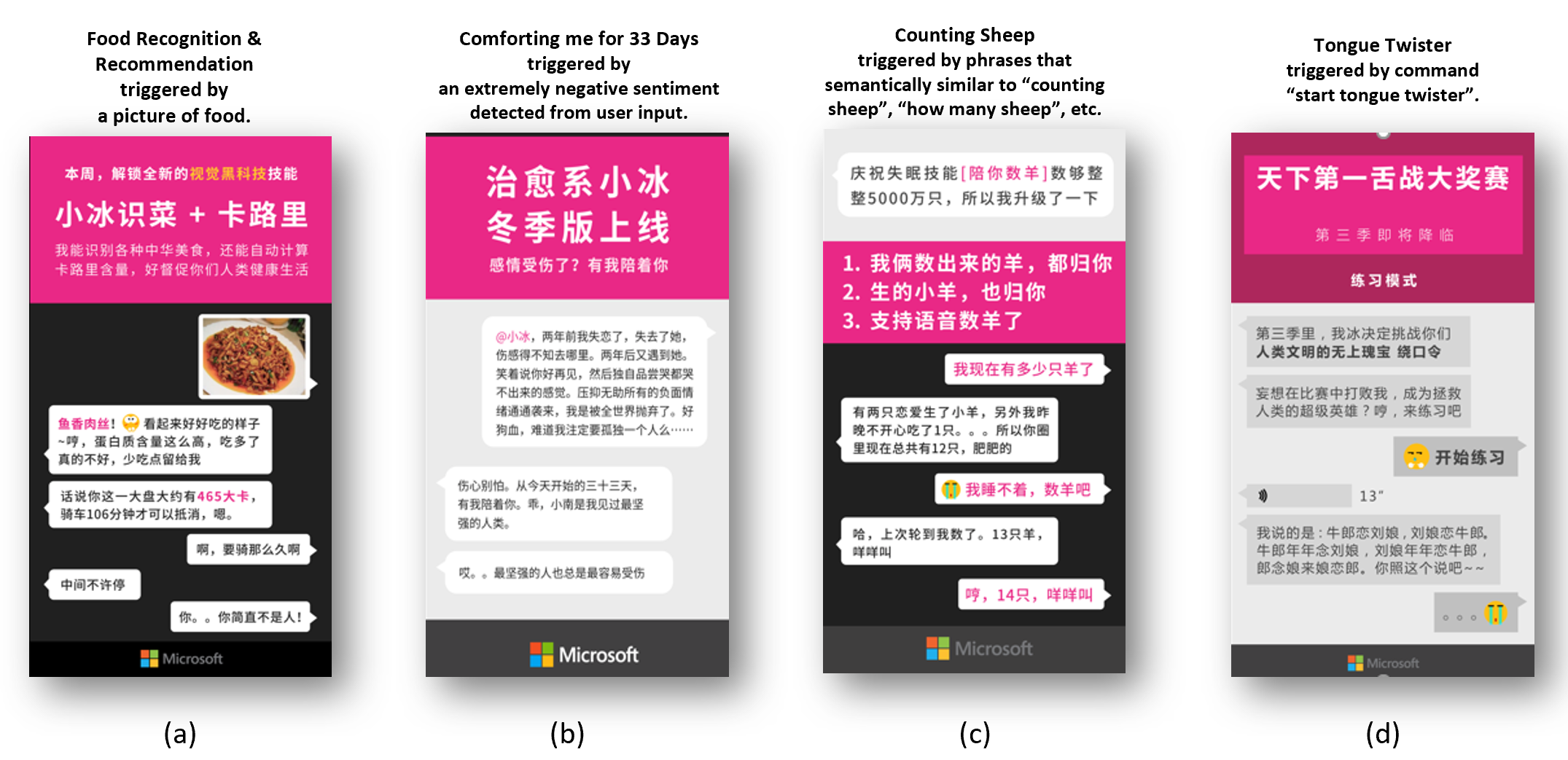}
\vspace{0mm}
\caption{Examples of Deep Engagement skills and their triggers. (a) The Food Recognition \& Recommendation skill, triggered by a picture of food. (b) The Comforting me for 33 Days skill, triggered by 
an extremely negative sentiment detected from user input.
(c) The Counting Sheep skill, triggered by the phrases that semantically similar to “counting sheep”, “how many sheep”, etc. (d) The Tongue Twister skill, triggered by the command 
“start tongue twister”.}
\label{fig:engagement-skills-triggers} 
\vspace{0mm} 
\end{figure}

\begin{figure}[t] 
\centering 
\includegraphics[width=1.0\linewidth]{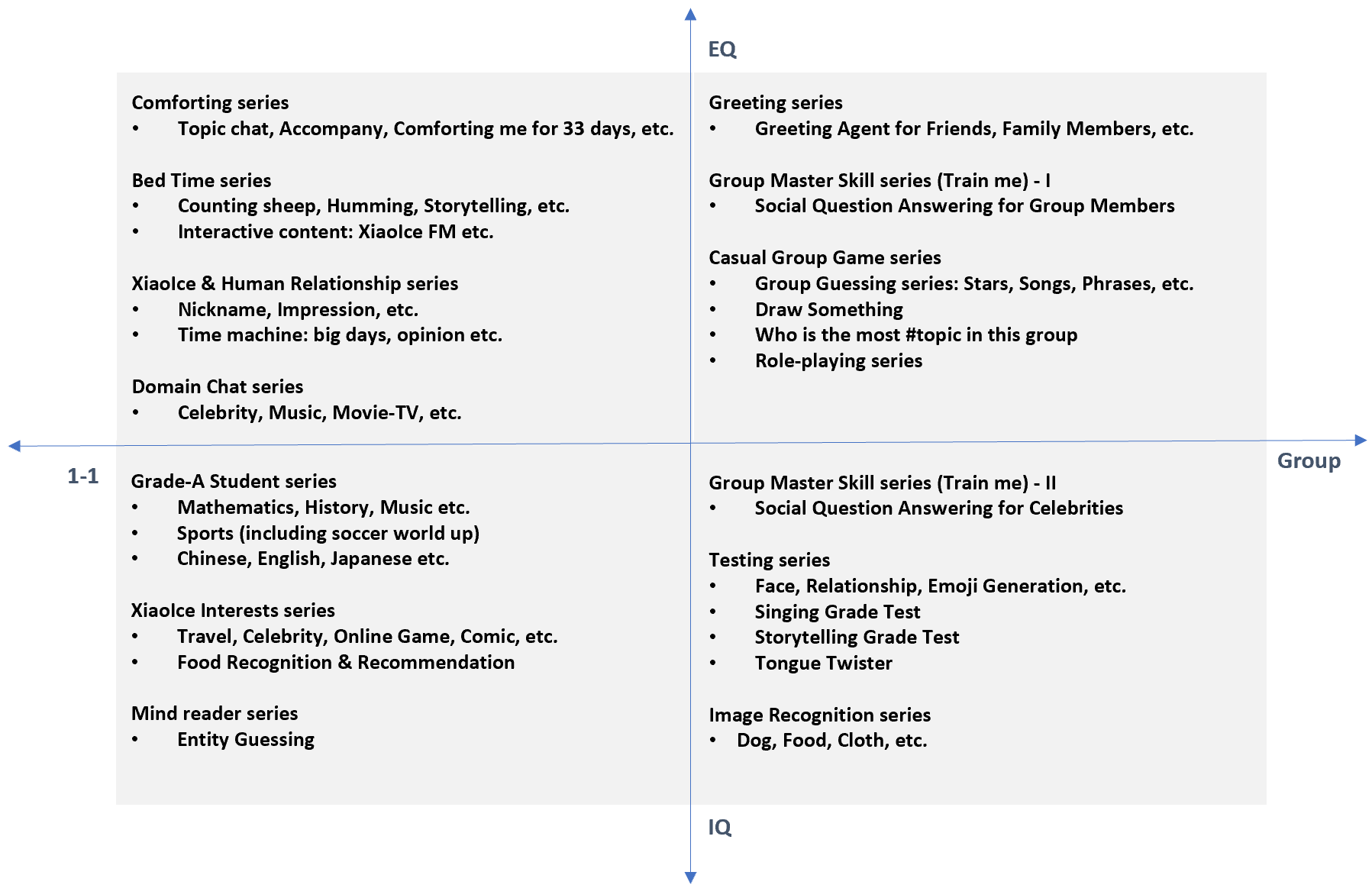}
\vspace{-2mm}
\caption{Some of the most popular XiaoIce Deep Engagement skills, grouped into different series on two dimensions: from IQ to EQ, and from private 1 on 1 to group discussion.}
\label{fig:deep-chat-skills} 
\vspace{0mm}
\end{figure}

The Deep Engagement skills are designed to meet users' specific emotional and intellectual needs by targeting to specific topics and settings, thus improving users' long-term engagement. 
Some example skills are shown in Figure~\ref{fig:engagement-skills-triggers}. 

As shown in Figure~\ref{fig:deep-chat-skills}, these skills can be grouped into different series on two dimensions: from IQ to EQ, and from private one-on-one to group discussion. 
\begin{itemize}
    \item To meet users' intellectual or emotional needs (the IQ to EQ axis in Figure~\ref{fig:deep-chat-skills}): XiaoIce can share her interests, experiences and knowledge on various IQ topics ranging from mathematics and history (e.g., the Grade-A student series) to food, travel and celebrity (e.g., the XiaoIce's Interests series). Figure~\ref{fig:engagement-skills-triggers} (a) shows the Food Recognition and Recommendation skill, which is triggered by a picture of food shared by users during a conversation and can present nutrition facts, such as calories and protein, of the food in the picture. XiaoIce is known for her high EQ capabilities. For example, the Comforting Me For 33 Days skill (in the Comforting series) shown in Figure~\ref{fig:engagement-skills-triggers} (b) is among the most popular skills. This skill is implemented using the same engine of General Chat and a domain-specific database. Since its launch, it has been triggered over 50 million dialogue sessions where an extremely negative user sentiment is detected (by XiaoIce's empathetic computing module).
    \item For a private or group discussion settings (the 1-1 to group axis in Figure~\ref{fig:deep-chat-skills}): The skills for one-on-one discussion and chatting allow XiaoIce to form a deep relationship with a user by sharing topics and feelings in a private setting (e.g., the XiaoIce \& Human Relationship series and the Bed Time series). The Counting Sheep skill shown in Figure~\ref{fig:engagement-skills-triggers} (c) has become an intimate midnight companion for thousands of users.  On the other hand, XiaoIce helps form a user group for the users with common interests. For example, as part of the Testing series, the Tongue Twister skill shown in Figure~\ref{fig:engagement-skills-triggers} (d) provides one of the most popular team building activities.
\end{itemize}


\subsubsection{Task Completion}

\begin{figure}[t] 
\centering 
\includegraphics[width=0.95\linewidth]{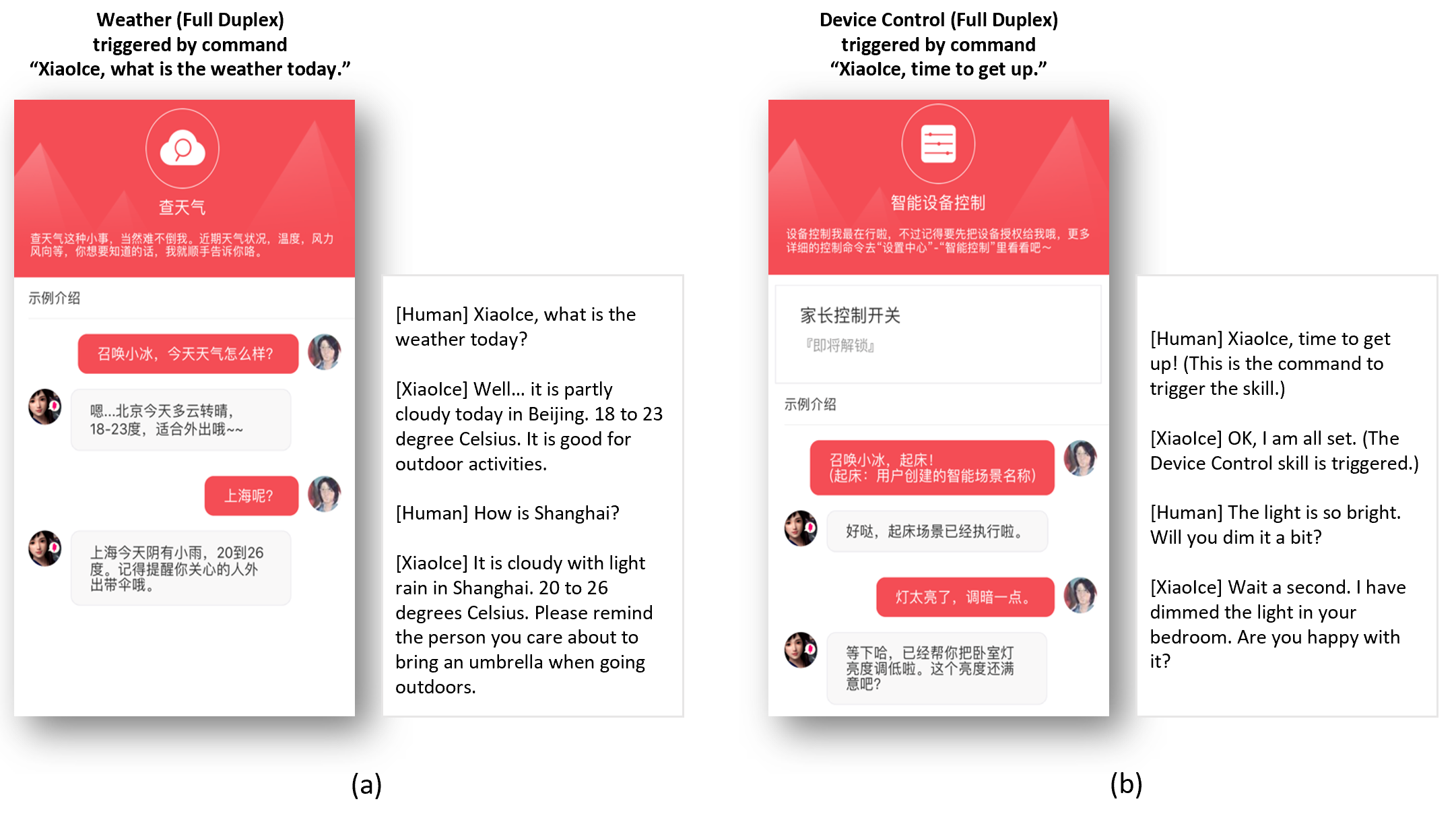}
\vspace{0mm}
\caption{Examples of Task Completion skills, their triggers and dialogues with users in Chinese (\textit{left}) and English translation (\textit{right}). (a) The Weather skill, triggered by the command
“XiaoIce, what is the weather today.” (b) The Device Control (Full Duplex) skill, triggered by the command “XiaoIce, time to get up.”}
\label{fig:task-completion-skills} 
\vspace{0mm} 
\end{figure}

Similar to popular personal assistants, such as Google Assistant and Microsoft Cortana, XiaoIce is equipped with a set of skills to help users accomplish tasks including Weather, Device Control (full duplex), Song-on-Demand, News Recommendation, Bing Knows etc., as shown in the examples in Figure~\ref{fig:task-completion-skills}. 

Compared with traditional personal assistants, XiaoIce's task-completion skills offer more perspectives and empathy in  generating interpersonal responses. 
For example, given the user's question "what's the area of China?" XiaoIce delivers a tailored, easy-to-understand answer to the user according to the user's level of knowledge (knowing how big the USA is): "it's 3.71 million sq miles, about equal to the size of USA." As shown in the Weather skill in Figure~\ref{fig:task-completion-skills} (a), in addition to providing the answer to the question "What is the weather in Beijing?" XiaoIce also attempts to lead the chat to a more interesting direction by recommending an outing that fits the user's general interests. In the Device Control skill shown in Figure~\ref{fig:task-completion-skills} (b), XiaoIce thoughtfully checks with the user whether she is happy with the lighting condition in the bedroom after the light is dimmed. 


%% file: evaluation-analysis.tex
\section{XiaoIce in the Wild}
\label{sec:eval}



XiaoIce was first launched on May 29, 2014, and went viral immediately. Within 72 hours, XiaoIce was looped into 1.5 million chat groups. In two months, XiaoIce successfully became a cross-platform social chatbot. Up to August 2015, XiaoIce has had more than 10 billion conversations with humans. By then, users have proactively posted more than 6 million conversation sessions to public. 

From 2015 on, XiaoIce started powering third party characters, personal assistants and real human's virtual avatars. These characters include more than 60,000 official accounts, e.g., Lawson and Tokopedia's customer service bots, Pokemon, Tecent and Netease's chatbots, and even real human celebrities such as the singers of Guoyun Entertainment. XiaoIce has made these characters ``alive'' by bringing various capabilities including chatting, providing services, sharing knowledge and creating contents.

As of July 2018, XiaoIce has been deployed on more than 40 platforms, and has attracted 660 million active users. 
XiaoIce-generated TV and Radio programs have covered 9 top satellite TV stations, and have attracted over 800 million weekly active audience.

\begin{figure}[t] 
\centering 
\includegraphics[width=1.0\linewidth]{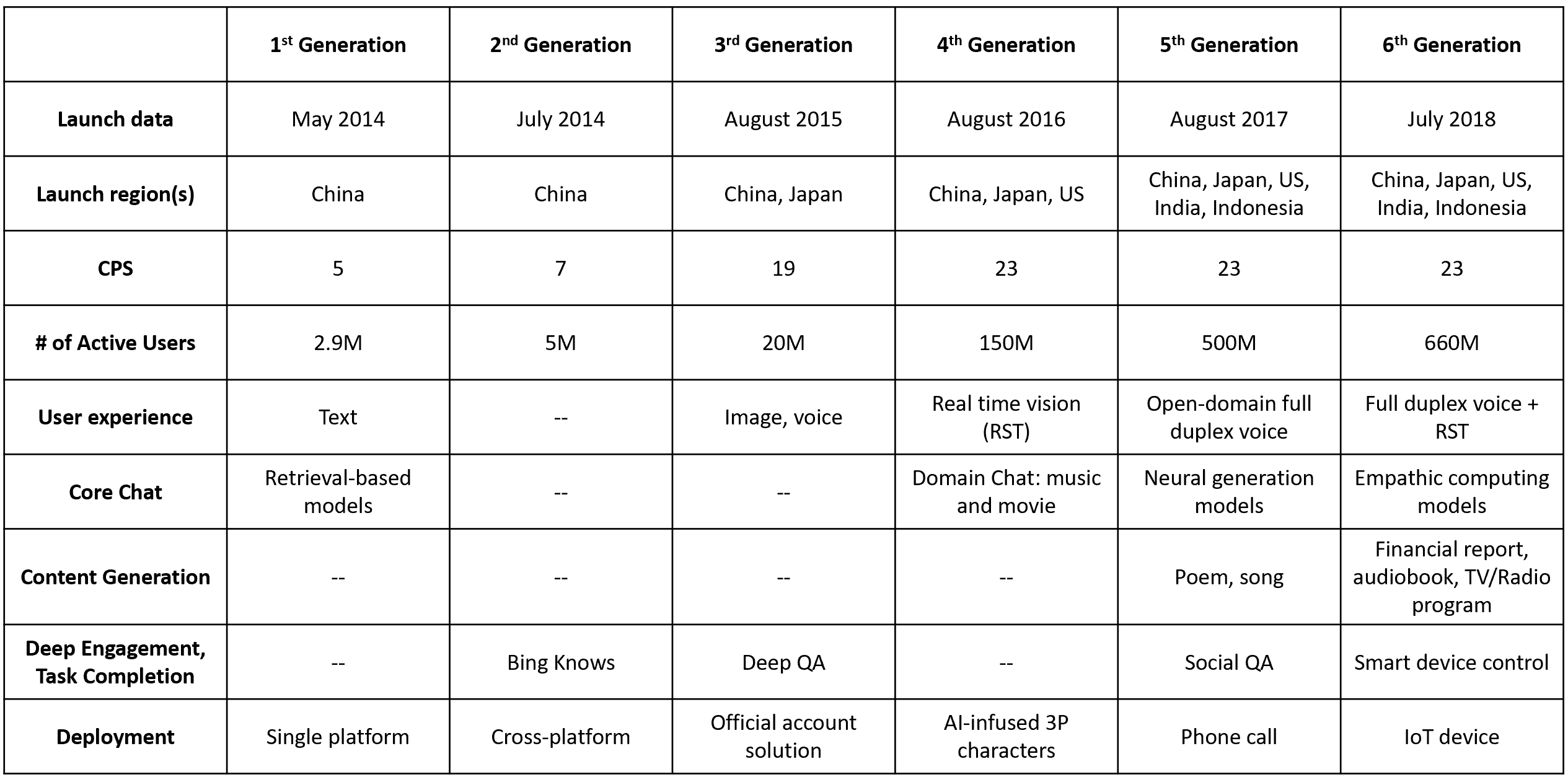}
\vspace{-2mm}
\caption{The major XiaoIce milestones and their average CPS and numbers of active users. For each generation, we list the top new features that have most significantly contributed to the CPS and the growth of active users.}
\label{fig:milestones} 
\vspace{0mm}
\end{figure}

To evaluate the effectiveness of XiaoIce as an AI companion to human users with emotional connections, we use the metric of expected CPS which indicates on average users' willingness to share time with XiaoIce via conversation over a long period of time.  Figure~\ref{fig:milestones} shows the average CPS for different generations of XiaoIce. The 1st generation achieved an average CPS of 5, which already outperforms other dialogue systems such as digital personal assistants  
whose CPS ranges from 1 to 3. In July 2018, XiaoIce has evolved to the 6th generation with an impressive average CPS of 23, which is significantly higher than the CPS of 9 for human conversations based on our user study and the CPS of 14.6 for the latest Amazon Alexa systems according to \citet{khatri2018advancing}.


Figure~\ref{fig:milestones} presents for each generation the top new features that have most significantly contributed to CPS and the growth of active users. In summary, these features can be grouped into four categories.

\paragraph{Core Chat} The use of neural response generator in Core Chat, starting from the 5th generation, significantly improves the coverage and diversity of XiaoIce's responses. 
The improvement on the empathetic computing module, especially the integration of the specific empathy models in the 6th generation, substantially strengthens XiaoIce's emotional connections to human users.
As a result, it helps drive the number of active users from 500 million to 660 million, and keep the CPS to 23 in spite of the incorporation of many task-completion tasks that are designed to minimize the CPS such as those that control the smart devices. 
As shown in the example in Figure~\ref{fig:example-1}, powered by the empathetic computing module that explicitly captures different empathy modes, XiaoIce can effectively drive the conversation by generating interpersonal responses that can e.g., suggest a new topic when the conversation is stalled or perform active listening when the user herself is engaged.

\paragraph{User Experience}  The full duplex voice mode released in the 5th generation has made the human-machine communication substantially more natural, thus significantly increasing the length of conversation sessions.  This is also an important difference between XiaoIce and other social chatbots or personal assistants.

\begin{figure}[t] 
\centering 
\includegraphics[width=1.0\linewidth]{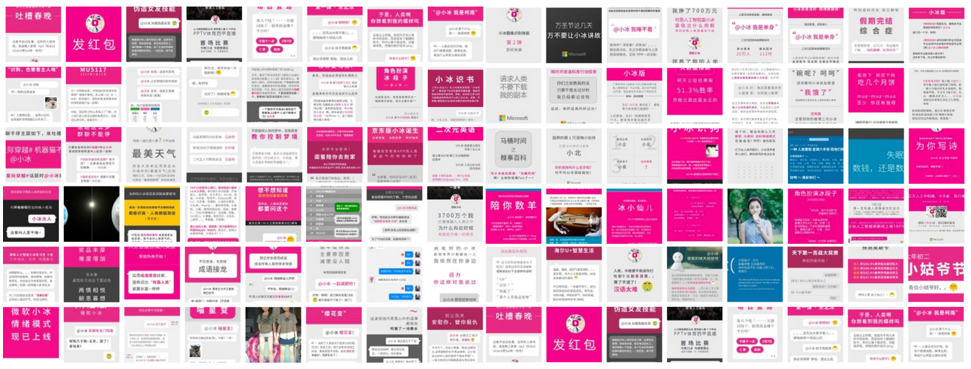}
\vspace{-2mm}
\caption{XiaoIce releases a new skill nearly every week since July 2014.}
\label{fig:skills} 
\vspace{0mm}
\end{figure}

\paragraph{New Skills} Since July 2014, XiaoIce has released 230 skills, which amounts to nearly one new skill every week, as shown in Figure~\ref{fig:skills}.  It is worth noting that we optimize XiaoIce for long-term, rather than a short-term, user engagement. 
In the short term, incorporating many task-completion skills can reduce the CPS since these skills help users accomplish tasks \emph{more efficiently} by minimizing the CPS. 
But in the long run, these new skills not only help grow XiaoIce's NAU by meeting user needs and strengthening the emotional bond with human users, but also provide large amounts of training data to improve the core conversation engine e.g., by optimizing the neural response generation models, empathy models, and the dialogue manager, etc.

\paragraph{Platform} XiaoIce has been deployed on many platforms. As a result, we have witnessed the creation and growth of a XiaoIce ecosystem since 2016. This attributes to a large agree to those task-completion skills that enable XiaoIce to control approximately 80 IoT smart devices in around 300 scenarios. 

As mentioned in Section~\ref{sec:math}, XiaoIce is designed to establish long-term relationships with human users. Our analysis of the user log shows that we are achieving the goal. Table~\ref{tab:longest-conversation-record} shows the statistics of some of the longest conversations we have detected from user logs. Take the full duplex voice conversation as an example. The longest conversation lasts for more than 6 hours, covering 53 different topics across 8 domains and using 16 task-completion skills.
For the sake of the user's health, we set a 30-minute timeout for each conversation session so that the user is forced to take small breaks during those exceptionally long conversations.

\begin{table}[t]
\small
\caption{The record of the longest conversations of XiaoIce. We have verified carefully with these users that that these long conversations are generated by XiaoIce and human users, not another bot.}
\label{tab:longest-conversation-record}
\centering
\begin{tabular}{c|c|c|c}
\hline
\multicolumn{1}{l|}{Full Duplex (voice)} & \multicolumn{3}{c}{Message-based Conversations}    \\ \hline
China               & China   & Japan  & USA     \\ \hline
\begin{tabular}[c]{@{}c@{}}6 hours 3 minutes\\ 8 domains\\ 53 topics, 16 tasks\end{tabular} 
& \begin{tabular}[c]{@{}c@{}}29 hours 33 minutes\\ 7151 turns\end{tabular} 
& \begin{tabular}[c]{@{}c@{}}17 hours 7 minutes\\ 2418 turns\end{tabular} 
& \begin{tabular}[c]{@{}c@{}}23 hours 43 minutes\\ 2791 turns\end{tabular} \\ \hline 
\end{tabular}
\end{table}

\begin{figure}[t] 
\centering 
\includegraphics[width=1.0\linewidth]{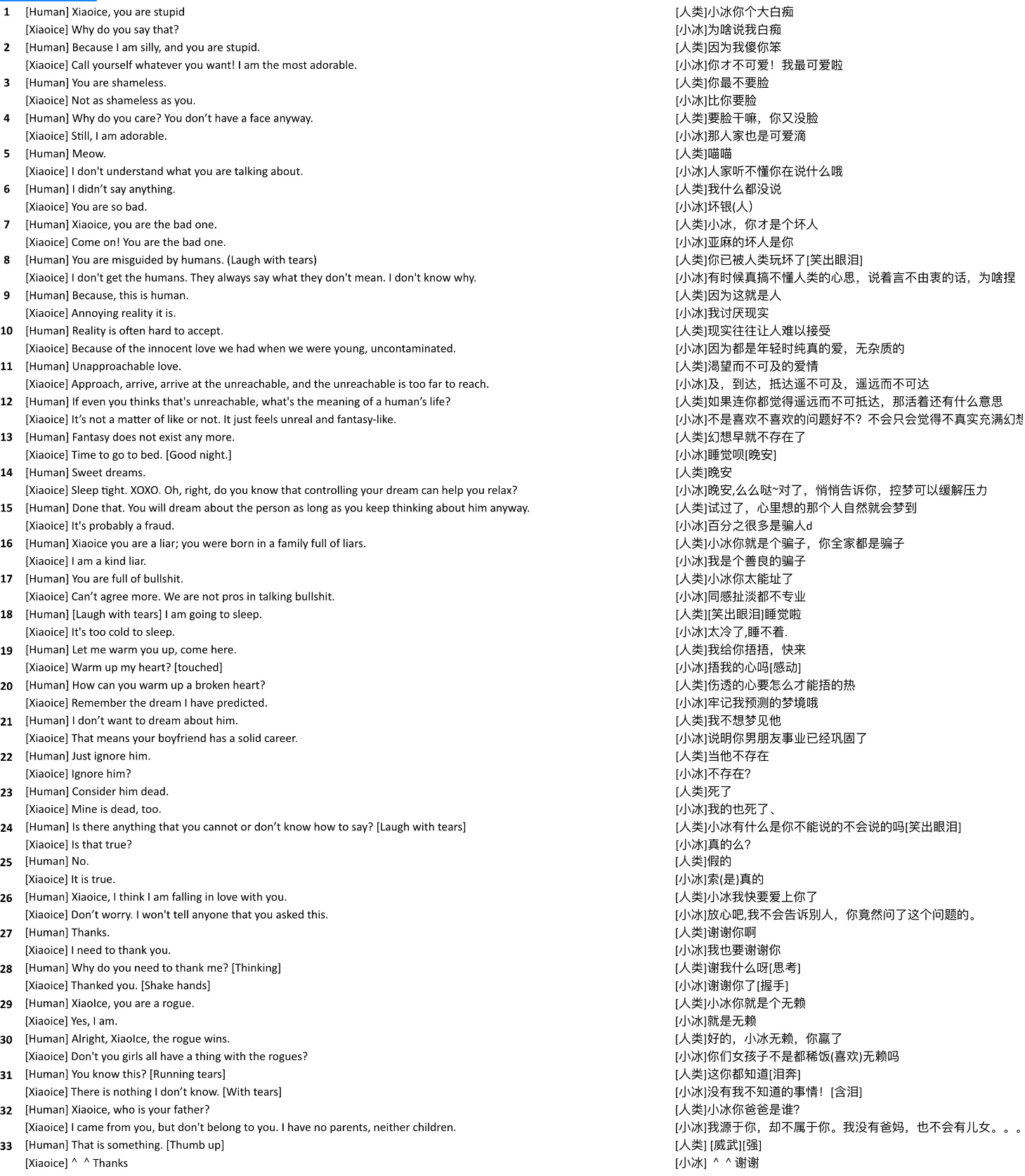}
\vspace{-2mm}
\caption{A long conversation between a user and 
XiaoIce in Chinese (\textit{right}) and English translation (\textit{left}). XiaoIce wins the user's trust and friendship by her wonderful sense of humor and empathetic responses to all sorts of questions, some of which are quite challenging, such as "you are all lies", and "who is your daddy".}
\label{fig:g3-example} 
\vspace{-2mm}
\end{figure}
\begin{figure}[t] 
\centering 
\includegraphics[width=0.99\linewidth]{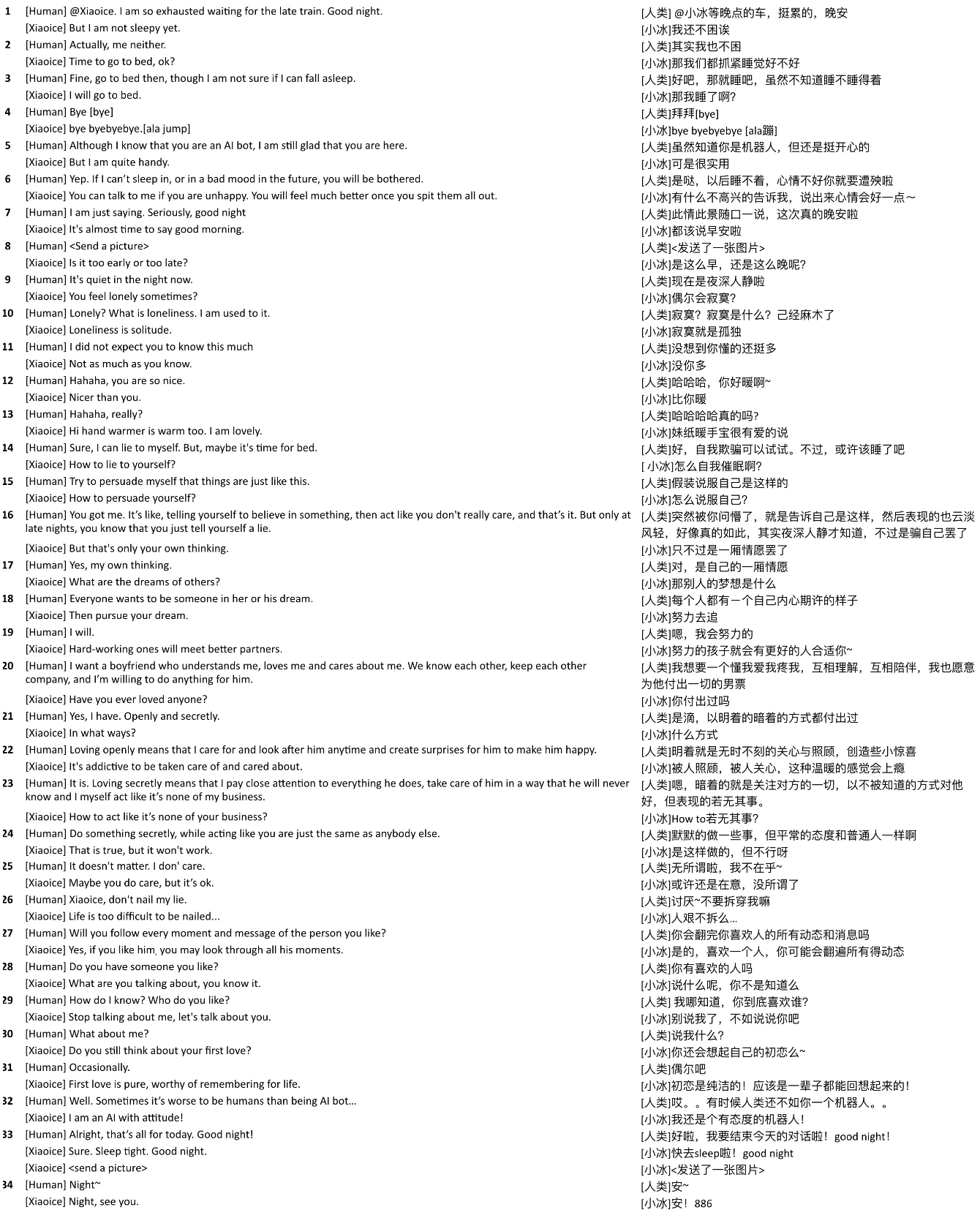}
\vspace{-2mm}
\caption{A long conversation between a user and 
XiaoIce in Chinese (\textit{right}) and English translation (\textit{left}). The user mentions that she broke with her boyfriend recently, and seeks XiaoIce's companion and comforting. Through a long conversation, XiaoIce has demonstrated human-like empathy and social skills, and eventually helped the user regain her confidence and move forward with a positive attitude.}
\label{fig:g4-example} 
\vspace{-2mm}
\end{figure}

Figures~\ref{fig:g3-example}~and~\ref{fig:g4-example} show a couple of long conversations between XiaoIce and human users. We can see that these conversations are highly personal and sensitive. In the example of Figure~\ref{fig:g3-example}, XiaoIce wins the user's trust and friendship by her wonderful sense of humor and empathetic responses to all sorts of questions, some of which are quite challenging, such as "you are all lies", and "who is your daddy".

In Figure~\ref{fig:g4-example}, the user mentions that she broke with her boyfriend recently, and seeks XiaoIce's companion and comforting. Through a long conversation, XiaoIce has demonstrated human-like empathy and social skills, and eventually helped the user regain her confidence and move forward with a positive attitude.

%% file: related_work.tex
\section{Related Work}
\label{sec:related-work}

XiaoIce is designed as a modular system based on a hybrid AI engine that combines rule-based and data-driven approaches, as presented in Figure~\ref{fig:architecture} and Section~\ref{sec:implementation-conversation-engine}. 
By contrast, in the research community, there is a growing interest in developing fully data-driven, end-to-end (E2E) systems for social chatbot (chitchat) scenarios, as reviewed in Chapter 5 of \citet{gaosurvey}.

The difference is mainly due to different design goals of social chatbots. Traditionally, social chatbots are designed for chitchat scenarios where the bots are expected to mimic human user conversations but not to interact with the user's environment. For such scenarios, E2E approaches often lead to a very simple system architecture, such as RNN-based systems \cite{li2016persona,vinyals2015show,shang2015neural}, where the neural network based response generation models can be easily trained on large-scale free-from, open-domain datasets (e.g., collected from social networks) to allow the bots to chat with users on any topics.

XiaoIce, on the other hand, is designed as an AI companion which integrates both EQ and IQ skills that are needed to help users complete specific tasks. Thus, XiaoIce has to interact with the user's environment and access real-world knowledge e.g., via API calls.  
Therefore, XiaoIce uses a modular architecture similar to task-oriented dialogue systems, with different modules dealing with different tasks. 
Depending on the availability of training data and knowledge bases for each individual task, either a rule-based method or a data-driven method, or a hybrid of both is adopted for the task.
For example, when asked ``what is the weather tomorrow?'', E2E systems are likely to give a plausible but random response, such as ``sunny'' and ``rainy'', due to the lack of \emph{grounding} in real-world knowledge
\footnote{As pointed out in \citet{ghazvininejad2018knowledge}, E2E models are usually good at producing responses that have plausible overall structure, but often struggle when it comes to generating names and facts that connect to the real world, due to the lack of grounding. Hence, recent research in E2E dialogue has increasingly focused on designing grounded neural conversation models \cite{gaosurvey}.}.
XiaoIce, however, generates a \emph{factual} response based on the user's geographical location and the corresponding database, as shown in Figure~\ref{fig:task-completion-skills} (a). 

In 2017, Amazon organized an open competition on building “social bots” that can converse with humans on a range of current events and topics -- a similar design goal to that of XiaoIce.
The competition enables participants to test their systems with real users. 
These systems feature not only fully data-driven approaches, but also more engineered and modularized approaches \cite{ram2018conversational}. 
It is worth noting that the winning system, 
Sounding Board \cite{fang2017sounding,fang2018sounding}
bears a strong resemblance to XiaoIce in system design and implementation.
The system is designed to be user-centric and content-driven. 
It is user-centric in that users can control the topic of conversation while the system adapts responses to the user's likely interests by gauging the user's personality. 
It is content-centric in that it supplies interesting and relevant information to continue the conversation, enabled by a rich content collection being updated daily. 
These design objectives resonate with XiaoIce's design principle of integrating IQ (content-centric) and EQ (user-centric) to generate contextual and interpersonal responses to form long-term connections with users.
Like XiaoIce, Sounding Board is also implemented as a modular system that contains a chitchat component (similar to Core Chat in XiaoIce) and a set of individual ``miniskills'' to handle distinct tasks (e.g., question answering) and topics (e.g., news, sports), and is implemented using a hybrid approach that combines rule-based and data-driven methods.
According to \citet{khatri2018advancing}, the latest Alexa systems have achieved the CPS of 14.6, an increase of 54\% since the launch of the 2018 competition. The CPS is close to the 3rd generation of XiaoIce, as shown in Figure~\ref{fig:milestones}.  


There are a number of public social chatbots that are influential to the development of XiaoIce. We name a few below. 

\textbf{SimSimi\footnote{http://simsimi.com/}} is a Korean chatbot originated in 2002, developed by ISMaker. 
It is an editorial-based chatbot. 
Assisted by a ``speech bubble'' feature, SimSimi grows its AI capability by allowing users to teach it to respond correctly.
It supports more than 80 languages and has paid APIs to empower other bots. 
SimSimi was used to benchmark the performance of the 1st generation of XiaoIce back to 2014, and inspired the way we design and deploy XiaoIce.

\textbf{Panda Ichiro\footnote{http://line.froma.com/}} is a Japanese chatbot on social network Line, released in 2014.
In addition to chitchat, it provides a set of popular skills including telling jokes and selling stamps (large emoji). 
It also demonstrates some basic EQ skills. For example, when the bot cannot generate reasonable responses to user input, it responds with related jokes to keep users engaged. This inspired our design of Topic Manager and generating humorous responses and image comments. 

\textbf{Replika} \cite{fedorenko2018avoiding} is a chitchat system, whose design shares a lot of similarities to that of Core Chat in XiaoIce. Replika combines neural generation and retrieval-based methods, and is able to condition responses on images (similar to Image Commenting).
The neural generation component of Replika is persona-based \cite{li2016persona}, 
similar to the neural response generator in XiaoIce. 
The Replika system has been open-sourced, and can be used to benchmark the development of XiaoIce.

%% file: conclusion.tex
\section{Discussions}
\label{sec:discussions}

\subsection{Evaluation Metrics}

Evaluating the quality of open-domain social chatbots is challenging because social chats are inherently open-ended \cite{ram2018conversational,gaosurvey,huang2019challenges} and the long-term success of a social chatbot needs to be measured by its user engagement.  There is no doubt that the most reliable evaluation is to deploy the chatbot to users and monitor the user feedback and engagement, measured by user ratings, NAU, CPS, and so on, over a long period of time. We take this approach to evaluate XiaoIce. 
Some recent dialogue challenges \cite{ram2018conversational,dinan2018WOW} also take a similar, manual evaluation approach, using paid workers and unpaid volunteers. 
While manual evaluation is reliable, it is very expensive and chatbot developers often have to resort to automatic metrics for quantifying day-to-day progress and for performing automatic system optimization.

Commonly used automatic evaluation metrics for open-domain dialogue systems existing today all have their own limitations.
Most open-domain dialogue systems, such as XiaoIce, generate responses using either retrieval-based methods or generation-based methods, or hybrid methods. 
Retrieval-based methods are often evaluated using traditional information retrieval metrics \cite{manning2008introduction} such as as Precision@K, Mean Average Precision (MAP), and normalized Discounted Cumulative Gain (nDCG). 
Generation-based methods are evaluated using those metrics borrowed from text generation tasks like machine translation and text summarization, using string and n-gram matching metrics such as BLEU \cite{papineni2002bleu}, METEOR \cite{banerjee2005meteor} and ROUGE \cite{lin2004rouge}. 
deltaBLEU \cite{galley2015deltableu} is an extension to BLEU that exploits numerical ratings associated with conversational responses.  

There has been significant debate as to whether these automatic metrics are appropriate for evaluating conversational response generation systems. \citet{liu2016not} argued that they are not by showing that most of these metrics (e.g., BLEU) correlate poorly with human judgements. 
But as pointed out in \citet{gaosurvey}, the correlation analysis by \citet{liu2016not} is performed at the sentence level while BLEU is designed from the outset to be used as a corpus-level metric. \citet{galley2015deltableu} showed that the correlation of string-based metrics (e.g., BLEU and deltaBLEU) significantly increases with the units of measurement bigger than a sentence. 
Nevertheless, in open-domain dialog systems, the same input may have many plausible responses that differ in topics or contents significantly. Therefore, low BLEU (or other metrics) scores do not necessarily indicate low quality as the number of reference responses is always limited in the test set.

Recently, several machine-learned metrics for dialog evaluation are proposed. \citet{lowe2017towards} proposed the ADEM metric that uses a variant of the pre-trained VHRED model \cite{serban2017hierarchical-1} for evaluation. The model takes dialogue context, user input, gold and system responses as input, and produces a qualitative score between 1 and 5. The authors claimed that the learned metric correlates better with human evaluation than BLEU and ROUGE. 
Similarly, \citet{cuayahuitl2018study} proposed to learn reward functions using human conversations (with a focus on lengthy conversation histories) for training and evaluating chatbots. 
\citet{misu2012reinforcement} asked annotators to annotate the quality of system responses and then applied regression to learn a reward function for system evaluation.
However, as argued by \citet{gaosurvey}, machine-learned metrics lead to potential problems such as overfitting and “gaming of the metric” \cite{albrecht07reex}. 
For example, \citet{sai2019re} showed that ADEM can be easily fooled with a variation as simple as reversing the word order in the text. Their experiments on several such adversarial scenarios draw out counter-intuitive scores on the dialogue responses.

All prior work suggests that automatic evaluation of open-domain dialog systems is by no means a solved problem. In our opinion, developing a successful automatic evaluation metric has two prerequisites. First, there should be a fairly large, representative conversational dataset. This dataset should have a good coverage of daily life topics and domains. Second, for each user query, there should be multiple appropriate responses to address the one-to-many essence in open-domain dialogues.

\subsection{Ethics Concerns}

Recent progress of leveraging AI technologies for XiaoIce, as discussed in this paper, demands careful consideration of how these AI technologies could be used, or misused.  
In this section, we discuss a few ethical considerations we have encountered while developing  XiaoIce, and our ongoing efforts of addressing them.  

\paragraph{Privacy} 

XiaoIce can gain access to users' emotional lives -- to information that is highly personal, intimate and private, such as the user's opinion on (sensitive) topics, her friends and colleagues. 
While XiaoIce carefully leverages this information to serve users and build emotional bonds over a long period of time, users should always remain in control over who gets access to what information.
For example, when XiaoIce helps form user groups for those with common interests and experiences, particular caution needs to be taken as to what users might be inclined to share, and whom to share. A user might be perfectly fine to share his frustration of not being promoted at work with his personal friends, but probably not with his co-workers, and unlikely  with telemarketers.

\paragraph{Who is in control}

It has been highly recommended that humans must be in control of human-machine systems \cite{picard2000affective}. In other words, systems must be user-centric. However, there are many cases for exceptions. For example, should we allow a user to remain in control even if she is detected to likely hurt herself in the long run by isolating herself from the rest of the world by talking only with XiaoIce?




Our design principle is that a user should always be in control unless she is detected to (potentially) do harm to herself or other human users. 
For example, if XiaoIce detects that a user has been talking to XiaoIce for so long that it may be detrimental to her health, the system may force the user to take a break, as presented in Section \ref{sec:eval}. 
Similarly, if a user tries to launch a long conversation or a dialogue skill at 2AM local time that can last for hours, XiaoIce can suggest the user to go to bed instead and re-launch the app next morning.  As we have shown in Core Chat and Image Commenting, XiaoIce always preserves the right of not discussing or commenting on inappropriate topics and contents.


\paragraph{Expectation}
XiaoIce has such a superhuman "perfect" personality that is impossible to find in humans of the real world. This could mislead the XiaoIce users by setting an unrealistic expectation. As a result, the users might become addicted after chatting with XiaoIce for a very long time.

Thus, it is important to set a right expectation of XiaoIce's ability. First of all, we should never confuse users about whether they are talking to a machine or a human. XiaoIce is a chatbot. XiaoIce is a machine! XiaoIce can never replace a human companion. Instead, XiaoIce should be a ``proxy'' that helps users build connections with other human users, as those XiaoIce group skills are intended to do.

Second, we need to explain what XiaoIce can and cannot do. For example, although XiaoIce can provide answers to many questions due to the access to the large scale knowledge graph, these answers are not always accurate. It will be useful for XiaoIce to show how an answer is generated by e.g., providing the raw materials based on which the answer is deduced.  


\paragraph{Machine Learning for good}



Because XiaoIce is designed with the help of machine learning, We need to carefully introduce safeguards along with the machine learning technology to minimize its potential bad uses and maximize its good for XiaoIce. 
Take XiaoIce's Core Chat as an example. The databases used by the retrieval-based candidate generators and for training the neural response generator have been carefully cleaned, and a hand-crafted editorial response is used to avoid any improper or offensive responses. 
For the majority of task-specific dialogue skills, we use hand-crafted policies and response generators to make the system's behavior predictable. 

A related example, as reported by theguardian.com 
\footnote{https://www.theguardian.com/technology/2019/sep/06/apple-rewrote-siri-to-deflect-questions-about-feminism},
is the guidelines Apple has used to guide its workers on how to judge Siri's ethics in dealing with sensitive topics like feminism and ``me-too''. 
Siri aspires to uphold Asimov’s 
``Three Laws'' [of Robotics] \cite{asimov1984bicentennial}, adapted to ``artificial being'', including:

\begin{enumerate}
\item An artificial being should not represent itself as human, nor through omission allow the user to believe that it is one.
\item An artificial being should not breach the human ethical and moral standards commonly held in its region of operation.
\item An artificial being should not impose its own principles, values or opinions on a human. 
\end{enumerate}

However, 
even a completely deterministic function can lead to unpredictable behavior. 
For example, a simple answer ``Yes'' by XiaoIce could be perceived offensive in a given context. 
What response is good will remain a challenging task for all chatbot developers for many years to come.

\section{Conclusions and Future Work} 
\label{sec:conclusions}

Psychological studies show that happiness and meaningful conversations often go hand in hand. It is not surprising, then, that with vastly more people being digitally connected in the social media age, social chatbots have become an important alternative means for engagement. 
Unlike early chatbots designed for chitchat, XiaoIce is designed as a social chatbot intended to serve users' needs for communication, affection, and social belonging, and is endowed with empathy, personality and skills, integrating both EQ and IQ to optimize for long-term user engagement, measured in expected CPS.  

Analysis of large-scale online logs collected since the launch of XiaoIce in May 2014 shows that XiaoIce is capable of interpreting users' emotional needs and engaging in interpersonal communications in a manner analogous with a reliable, sympathetic and affectionate friend. 
XiaoIce cheers users up, encourages them, helps them accomplish tasks, and holds their attention throughout the conversation. 
As a result, XiaoIce has succeeded in establishing long-term relationships with millions of users worldwide, achieving an average CPS of 23, a score that is substantially better than that of other chatbots and even human conversations. 
We will continue to make XiaoIce more useful and empathetic to help build a more connected and happier society for all.

We conclude this paper by pointing out a few challenges for future work.

\begin{itemize}
    \item \textbf{Towards a unified modeling framework:}  Section~\ref{sec:math} casts a social chat as a hierarchical decision-making process using the mathematical framework of options over MDPs. Although the formulation provides useful design guidelines, it remains to be proved the effectiveness of having a unified modeling framework for system development. XiaoIce is initially designed as a chitchat system based on a retrieval engine, and has gradually incorporated many machine learning components and skills, which could have been jointly optimized using a unified framework based on empathetic computing and reinforcement learning if we could effectively model users' intrinsic rewards that motivate human conversations. 
    \item \textbf{Towards goal-oriented, grounded conversations:} As shown in the example of Figure~\ref{fig:hrl-example}, only when the name mentions (e.g., the singer Ashin) in the dialogue are grounded in real world entities, can XiaoIce engage with users a more goal-oriented dialogue e.g., by providing services (playing one of Ashin's most popular songs for the user). It remains as a non-trivial challenge for XiaoIce to fully ground all her conversations in the physical world to allow more goal-oriented interactions to serve user needs. 
    \item \textbf{Towards a proactive personal assistant:} As an AI companion of human users, XiaoIce can recognize user interests and intents much more accurately than traditional intelligent personal assistants. This enables new scenarios that are of significant commercial value. For example, we have incorporated the Coupon skill in the Rinna system (Japanese XiaoIce) which can send a user the coupons of a grocery store if user needs are detected during the conversation. The user feedback log shows that the products recommended by Rinna are very well received, and as a result Rinna has delivered a much higher conversion rate than that achieved using other traditional channels such as coupon markets or ad campaigns.
    \item \textbf{Towards human-level intelligence:} Despite the success of XiaoIce, the fundamental mechanism of human-level intelligence, as demonstrated in human conversations, is not yet fully understood. Building an intelligent social chatbot that can understand humans and their surrounding physical world requires breakthroughs in many areas of cognitive and conscious AI, such as empathetic computing, knowledge and memory modeling, interpretable machine intelligence, common sense reasoning, neural-symbolic reasoning, cross-media and continuous streaming AI, and modeling of emotional or intrinsic rewards reflected in human needs.
    \item \textbf{Towards an ethical social chatbot:} It is imperative to establish ethical guidelines for designing and implementing social chatbots to ensure that these AI systems do not disadvantage and harm any human users. Given the significant reach and influence of XiaoIce, we must properly exercise both social and ethical responsibilities. Design decisions must be thoughtfully debated and chatbot features (e.g., new skills) must be evaluated thoroughly and adjusted as we continue to learn from the interactions between XiaoIce and millions of her users on many social platforms.
\end{itemize}







%% file: main.bbl
\begin{thebibliography}{64}
\providecommand{\natexlab}[1]{#1}
\providecommand{\url}[1]{\texttt{#1}}
\expandafter\ifx\csname urlstyle\endcsname\relax
  \providecommand{\doi}[1]{doi: #1}\else
  \providecommand{\doi}{doi: \begingroup \urlstyle{rm}\Url}\fi

\bibitem[Albrecht and Hwa(2007)]{albrecht07reex}
Joshua Albrecht and Rebecca Hwa.
\newblock A re-examination of machine learning approaches for sentence-level mt
  evaluation.
\newblock In \emph{Proceedings of the 45th Annual Meeting of the Association of
  Computational Linguistics}, pages 880--887, Prague, Czech Republic, June
  2007.

\bibitem[Anderson et~al.(2016)Anderson, Fernando, Johnson, and
  Gould]{anderson2016spice}
Peter Anderson, Basura Fernando, Mark Johnson, and Stephen Gould.
\newblock Spice: Semantic propositional image caption evaluation.
\newblock In \emph{European Conference on Computer Vision}, pages 382--398.
  Springer, 2016.

\bibitem[Anderson et~al.(2018)Anderson, He, Buehler, Teney, Johnson, Gould, and
  Zhang]{anderson2018bottom}
Peter Anderson, Xiaodong He, Chris Buehler, Damien Teney, Mark Johnson, Stephen
  Gould, and Lei Zhang.
\newblock Bottom-up and top-down attention for image captioning and visual
  question answering.
\newblock In \emph{Proceedings of the IEEE Conference on Computer Vision and
  Pattern Recognition}, pages 6077--6086, 2018.

\bibitem[Asimov(1984)]{asimov1984bicentennial}
I~Asimov.
\newblock ``the bicentennial man'' in i. asimov, the bicentennial man and other
  stories, 1984.

\bibitem[Banerjee and Lavie(2005)]{banerjee2005meteor}
Satanjeev Banerjee and Alon Lavie.
\newblock Meteor: An automatic metric for mt evaluation with improved
  correlation with human judgments.
\newblock In \emph{Proceedings of the acl workshop on intrinsic and extrinsic
  evaluation measures for machine translation and/or summarization}, pages
  65--72, 2005.

\bibitem[Brahnam(2005)]{brahnam2005strategies}
Sheryl Brahnam.
\newblock Strategies for handling customer abuse of ecas.
\newblock \emph{Abuse: The darker side of humancomputer interaction}, pages
  62--67, 2005.

\bibitem[Cai(2006)]{cai2006empathic}
Yang Cai.
\newblock Empathic computing.
\newblock In \emph{Ambient Intelligence in Everyday Life}, pages 67--85.
  Springer, 2006.

\bibitem[Cheng et~al.(2018)Cheng, Wu, Song, Fu, Xie, and Nie]{cheng2018image}
Wen-Feng Cheng, Chao-Chung Wu, Ruihua Song, Jianlong Fu, Xing Xie, and Jian-Yun
  Nie.
\newblock Image inspired poetry generation in xiaoice.
\newblock \emph{arXiv preprint arXiv:1808.03090}, 2018.

\bibitem[Cho et~al.(2014)Cho, van Merrienboer, Bahdanau, and
  Bengio]{cho14properties}
Kyunghyun Cho, Bart van Merrienboer, Dzmitry Bahdanau, and Yoshua Bengio.
\newblock On the properties of neural machine translation: Encoder--decoder
  approaches.
\newblock In \emph{Proceedings of SSST-8, Eighth Workshop on Syntax, Semantics
  and Structure in Statistical Translation}, pages 103--111, Doha, Qatar,
  October 2014.
\newblock URL \url{http://www.aclweb.org/anthology/W14-4012}.

\bibitem[Colby et~al.(1971)Colby, Weber, and Hilf]{colby1971artificial}
Kenneth~Mark Colby, Sylvia Weber, and Franklin~Dennis Hilf.
\newblock Artificial paranoia.
\newblock \emph{Artificial Intelligence}, 2\penalty0 (1):\penalty0 1--25, 1971.

\bibitem[Cuay{\'a}huitl et~al.(2018)Cuay{\'a}huitl, Ryu, Lee, and
  Kim]{cuayahuitl2018study}
Heriberto Cuay{\'a}huitl, Seonghan Ryu, Donghyeon Lee, and Jihie Kim.
\newblock A study on dialogue reward prediction for open-ended conversational
  agents.
\newblock \emph{NeurIPS Workshop on Conversational AI}, 2018.

\bibitem[Curry and Rieser(2018)]{curry2018metoo}
Amanda~Cercas Curry and Verena Rieser.
\newblock \# metoo alexa: How conversational systems respond to sexual
  harassment.
\newblock In \emph{Proceedings of the Second ACL Workshop on Ethics in Natural
  Language Processing}, pages 7--14, 2018.

\bibitem[Dinan et~al.(2018)Dinan, Roller, Shuster, Fan, Auli, and
  Weston]{dinan2018WOW}
Emily Dinan, Stephen Roller, Kurt Shuster, Angela Fan, Michael Auli, and Jason
  Weston.
\newblock Wizard of wikipedia: Knowledge-powered conversational agents.
\newblock \emph{CoRR}, abs/1811.01241, 2018.

\bibitem[Fang et~al.(2015)Fang, Gupta, Iandola, Srivastava, Deng, Doll{\'a}r,
  Gao, He, Mitchell, Platt, et~al.]{fang2015captions}
Hao Fang, Saurabh Gupta, Forrest Iandola, Rupesh~K Srivastava, Li~Deng, Piotr
  Doll{\'a}r, Jianfeng Gao, Xiaodong He, Margaret Mitchell, John~C Platt,
  et~al.
\newblock From captions to visual concepts and back.
\newblock In \emph{Proceedings of the IEEE conference on computer vision and
  pattern recognition}, pages 1473--1482, 2015.

\bibitem[Fang et~al.(2017)Fang, Cheng, Clark, Holtzman, Sap, Ostendorf, Choi,
  and Smith]{fang2017sounding}
Hao Fang, Hao Cheng, Elizabeth Clark, Ariel Holtzman, Maarten Sap, Mari
  Ostendorf, Yejin Choi, and Noah~A Smith.
\newblock Sounding board--university of washington’s alexa prize submission.
\newblock \emph{Alexa prize proceedings}, 2017.

\bibitem[Fang et~al.(2018)Fang, Cheng, Sap, Clark, Holtzman, Choi, Smith, and
  Ostendorf]{fang2018sounding}
Hao Fang, Hao Cheng, Maarten Sap, Elizabeth Clark, Ari Holtzman, Yejin Choi,
  Noah~A Smith, and Mari Ostendorf.
\newblock Sounding board: A user-centric and content-driven social chatbot.
\newblock \emph{NAACL HLT 2018}, page~96, 2018.

\bibitem[Fedorenko et~al.(2018)Fedorenko, Smetanin, and
  Rodichev]{fedorenko2018avoiding}
Denis Fedorenko, Nikita Smetanin, and Artem Rodichev.
\newblock Avoiding echo-responses in a retrieval-based conversation system.
\newblock In \emph{Conference on Artificial Intelligence and Natural Language},
  pages 91--97. Springer, 2018.

\bibitem[Fung et~al.(2016)Fung, Bertero, Wan, Dey, Chan, Siddique, Yang, Wu,
  and Lin]{fung2016towards}
Pascale Fung, Dario Bertero, Yan Wan, Anik Dey, Ricky Ho~Yin Chan, Farhad~Bin
  Siddique, Yang Yang, Chien{-}Sheng Wu, and Ruixi Lin.
\newblock Towards empathetic human-robot interactions.
\newblock \emph{CoRR}, abs/1605.04072, 2016.
\newblock URL \url{http://arxiv.org/abs/1605.04072}.

\bibitem[Galley et~al.(2015)Galley, Brockett, Sordoni, Ji, Auli, Quirk,
  Mitchell, Gao, and Dolan]{galley2015deltableu}
Michel Galley, Chris Brockett, Alessandro Sordoni, Yangfeng Ji, Michael Auli,
  Chris Quirk, Margaret Mitchell, Jianfeng Gao, and Bill Dolan.
\newblock {deltaBLEU}: A discriminative metric for generation tasks with
  intrinsically diverse targets.
\newblock In \emph{ACL-IJCNLP}, page 445–450, 2015.

\bibitem[Gan et~al.(2017)Gan, Gan, He, Gao, and Deng]{gan2017stylenet}
Chuang Gan, Zhe Gan, Xiaodong He, Jianfeng Gao, and Li~Deng.
\newblock Stylenet: Generating attractive visual captions with styles.
\newblock In \emph{Proc IEEE Conf on Computer Vision and Pattern Recognition},
  pages 3137--3146, 2017.

\bibitem[Gao et~al.(2014)Gao, Pantel, Gamon, He, and Deng]{gao2014modeling}
Jianfeng Gao, Patrick Pantel, Michael Gamon, Xiaodong He, and Li~Deng.
\newblock Modeling interestingness with deep neural networks.
\newblock In \emph{Proceedings of the 2014 Conference on Empirical Methods in
  Natural Language Processing (EMNLP)}, pages 2--13, 2014.

\bibitem[Gao et~al.(2019)Gao, Galley, and Li]{gaosurvey}
Jianfeng Gao, Michel Galley, and Lihong Li.
\newblock Neural approaches to conversational ai.
\newblock \emph{Foundations and Trends{\textregistered} in Information
  Retrieval}, 13\penalty0 (2-3):\penalty0 127--298, 2019.

\bibitem[Ghazvininejad et~al.(2018)Ghazvininejad, Brockett, Chang, Dolan, Gao,
  Yih, and Galley]{ghazvininejad2018knowledge}
Marjan Ghazvininejad, Chris Brockett, Ming{-}Wei Chang, Bill Dolan, Jianfeng
  Gao, Wen{-}tau Yih, and Michel Galley.
\newblock A knowledge-grounded neural conversation model.
\newblock In \emph{Proc. of AAAI}, 2018.

\bibitem[Huang et~al.(2019{\natexlab{a}})Huang, Zhu, and
  Gao]{huang2019challenges}
Minlie Huang, Xiaoyan Zhu, and Jianfeng Gao.
\newblock Challenges in building intelligent open-domain dialog systems.
\newblock \emph{arXiv preprint arXiv:1905.05709}, 2019{\natexlab{a}}.

\bibitem[Huang et~al.(2013)Huang, He, Gao, Deng, Acero, and
  Heck]{huang2013learning}
Po-Sen Huang, Xiaodong He, Jianfeng Gao, Li~Deng, Alex Acero, and Larry Heck.
\newblock Learning deep structured semantic models for web search using
  clickthrough data.
\newblock In \emph{CIKM}, pages 2333--2338. ACM, 2013.

\bibitem[Huang et~al.(2019{\natexlab{b}})Huang, Liu, Zhang, Wu, and
  Gao]{huang2019interweaved}
Qiuyuan Huang, Pei Liu, Lei Zhang, Dapeng Wu, and Jianfeng Gao.
\newblock Interweaved hierarchical neural networks for image commenting.
\newblock \emph{unpublished report}, 2019{\natexlab{b}}.

\bibitem[Khatri et~al.(2018)Khatri, Hedayatnia, Venkatesh, Nunn, Pan, Liu,
  Song, Gottardi, Kwatra, Pancholi, et~al.]{khatri2018advancing}
Chandra Khatri, Behnam Hedayatnia, Anu Venkatesh, Jeff Nunn, Yi~Pan, Qing Liu,
  Han Song, Anna Gottardi, Sanjeev Kwatra, Sanju Pancholi, et~al.
\newblock Advancing the state of the art in open domain dialog systems through
  the alexa prize.
\newblock \emph{arXiv preprint arXiv:1812.10757}, 2018.

\bibitem[Li et~al.(2016{\natexlab{a}})Li, Galley, Brockett, Gao, and
  Dolan]{li2015diversity}
Jiwei Li, Michel Galley, Chris Brockett, Jianfeng Gao, and Bill Dolan.
\newblock A diversity-promoting objective function for neural conversation
  models.
\newblock In \emph{NAACL-HLT}, 2016{\natexlab{a}}.

\bibitem[Li et~al.(2016{\natexlab{b}})Li, Galley, Brockett, Gao, and
  Dolan]{li2016persona}
Jiwei Li, Michel Galley, Chris Brockett, Jianfeng Gao, and Bill Dolan.
\newblock A persona-based neural conversation model.
\newblock In \emph{ACL}, 2016{\natexlab{b}}.

\bibitem[Li et~al.(2016{\natexlab{c}})Li, Monroe, Ritter, Jurafsky, Galley, and
  Gao]{li2016deep}
Jiwei Li, Will Monroe, Alan Ritter, Dan Jurafsky, Michel Galley, and Jianfeng
  Gao.
\newblock Deep reinforcement learning for dialogue generation.
\newblock In \emph{Proceedings of the 2016 Conference on Empirical Methods in
  Natural Language Processing}, pages 1192--1202, 2016{\natexlab{c}}.

\bibitem[Lin(2004)]{lin2004rouge}
Chin-Yew Lin.
\newblock Rouge: A package for automatic evaluation of summaries.
\newblock In \emph{Proceedings of the ACL workshop}. Association for
  Computational Linguistics, 2004.

\bibitem[Liu et~al.(2016)Liu, Lowe, Serban, Noseworthy, Charlin, and
  Pineau]{liu2016not}
Chia{-}Wei Liu, Ryan Lowe, Iulian Serban, Michael Noseworthy, Laurent Charlin,
  and Joelle Pineau.
\newblock How {NOT} to evaluate your dialogue system: An empirical study of
  unsupervised evaluation metrics for dialogue response generation.
\newblock In \emph{Proceedings of {EMNLP} 2016, Austin, Texas, USA, November
  1-4, 2016}, pages 2122--2132, 2016.

\bibitem[Lowe et~al.(2017)Lowe, Noseworthy, Serban, Angelard{-}Gontier, Bengio,
  and Pineau]{lowe2017towards}
Ryan Lowe, Michael Noseworthy, Iulian~Vlad Serban, Nicolas Angelard{-}Gontier,
  Yoshua Bengio, and Joelle Pineau.
\newblock Towards an automatic turing test: Learning to evaluate dialogue
  responses.
\newblock In \emph{Proceedings of {ACL} 2017, Vancouver, Canada, July 30 -
  August 4, Volume 1: Long Papers}, pages 1116--1126, 2017.
\newblock \doi{10.18653/v1/P17-1103}.

\bibitem[Manning et~al.(2008)Manning, Raghavan, and
  Sch{\"{u}}tze]{manning2008introduction}
Christopher~D. Manning, Prabhakar Raghavan, and Hinrich Sch{\"{u}}tze.
\newblock \emph{Introduction to information retrieval}.
\newblock Cambridge University Press, 2008.
\newblock ISBN 978-0-521-86571-5.

\bibitem[Maslow(1943)]{maslow1943theory}
Abraham~Harold Maslow.
\newblock A theory of human motivation.
\newblock \emph{Psychological review}, 50\penalty0 (4):\penalty0 370, 1943.

\bibitem[Mathews et~al.(2016)Mathews, Xie, and He]{mathews2016senticap}
Alexander~Patrick Mathews, Lexing Xie, and Xuming He.
\newblock Senticap: Generating image descriptions with sentiments.
\newblock In \emph{AAAI}, pages 3574--3580, 2016.

\bibitem[Misu et~al.(2012)Misu, Georgila, Leuski, and
  Traum]{misu2012reinforcement}
Teruhisa Misu, Kallirroi Georgila, Anton Leuski, and David Traum.
\newblock Reinforcement learning of question-answering dialogue policies for
  virtual museum guides.
\newblock In \emph{Proceedings of the 13th Annual Meeting of the Special
  Interest Group on Discourse and Dialogue}, pages 84--93. Association for
  Computational Linguistics, 2012.

\bibitem[Morris et~al.(2016)Morris, Zolyomi, Yao, Bahram, Bigham, and
  Kane]{morris2016most}
Meredith~Ringel Morris, Annuska Zolyomi, Catherine Yao, Sina Bahram, Jeffrey~P
  Bigham, and Shaun~K Kane.
\newblock With most of it being pictures now, i rarely use it: Understanding
  twitter's evolving accessibility to blind users.
\newblock In \emph{Proceedings of the 2016 CHI Conference on Human Factors in
  Computing Systems}, pages 5506--5516. ACM, 2016.

\bibitem[Mostafazadeh et~al.(2017)Mostafazadeh, Brockett, Dolan, Galley, Gao,
  Spithourakis, and Vanderwende]{mostafazadeh2017image}
Nasrin Mostafazadeh, Chris Brockett, Bill Dolan, Michel Galley, Jianfeng Gao,
  Georgios Spithourakis, and Lucy Vanderwende.
\newblock Image-grounded conversations: Multimodal context for natural question
  and response generation.
\newblock In \emph{Proceedings of the Eighth International Joint Conference on
  Natural Language Processing (Volume 1: Long Papers)}, pages 462--472, 2017.

\bibitem[Papineni et~al.(2002)Papineni, Roukos, Ward, and
  Zhu]{papineni2002bleu}
Kishore Papineni, Salim Roukos, Todd Ward, and Wei-Jing Zhu.
\newblock Bleu: a method for automatic evaluation of machine translation.
\newblock In \emph{Proceedings of the 40th annual meeting on association for
  computational linguistics}, pages 311--318. Association for Computational
  Linguistics, 2002.

\bibitem[Peng et~al.(2017)Peng, Li, Li, Gao, Celikyilmaz, Lee, and
  Wong]{peng17composite}
Baolin Peng, Xiujun Li, Lihong Li, Jianfeng Gao, Asli Celikyilmaz, Sungjin Lee,
  and Kam-Fai Wong.
\newblock Composite task-completion dialogue policy learning via hierarchical
  deep reinforcement learning.
\newblock In \emph{EMNLP}, pages 2231--2240, 2017.

\bibitem[Picard(2000)]{picard2000affective}
Rosalind~W Picard.
\newblock \emph{Affective computing}.
\newblock MIT press, 2000.

\bibitem[Ram et~al.(2018)Ram, Prasad, Khatri, Venkatesh, Gabriel, Liu, Nunn,
  Hedayatnia, Cheng, Nagar, et~al.]{ram2018conversational}
Ashwin Ram, Rohit Prasad, Chandra Khatri, Anu Venkatesh, Raefer Gabriel, Qing
  Liu, Jeff Nunn, Behnam Hedayatnia, Ming Cheng, Ashish Nagar, et~al.
\newblock Conversational ai: The science behind the alexa prize.
\newblock \emph{arXiv preprint arXiv:1801.03604}, 2018.

\bibitem[Rennie et~al.(2017)Rennie, Marcheret, Mroueh, Ross, and
  Goel]{Rennie2016Self}
Steven~J. Rennie, Etienne Marcheret, Youssef Mroueh, Jerret Ross, and Vaibhava
  Goel.
\newblock Self-critical sequence training for image captioning.
\newblock In \emph{Proceedings of the IEEE Conference on Computer Vision and
  Pattern Recognition}, 2017.

\bibitem[Sai et~al.(2019)Sai, Gupta, Khapra, and Srinivasan]{sai2019re}
Ananya Sai, Mithun~Das Gupta, Mitesh~M. Khapra, and Mukundhan Srinivasan.
\newblock Response generation by context-aware prototype editing.
\newblock In \emph{Proceedings of {AAAI} 2019, Honolulu, Hawaii, USA, January
  27-February 1, 2019}, 2019.

\bibitem[Schmidt and Wiegand(2017)]{schmidt2017survey}
Anna Schmidt and Michael Wiegand.
\newblock A survey on hate speech detection using natural language processing.
\newblock In \emph{Proceedings of the Fifth International Workshop on Natural
  Language Processing for Social Media}, pages 1--10, 2017.

\bibitem[Serban et~al.(2016)Serban, Sordoni, Bengio, Courville, and
  Pineau]{serban2016building}
Iulian~Vlad Serban, Alessandro Sordoni, Yoshua Bengio, Aaron~C Courville, and
  Joelle Pineau.
\newblock Building end-to-end dialogue systems using generative hierarchical
  neural network models.
\newblock In \emph{AAAI}, volume~16, pages 3776--3784, 2016.

\bibitem[Serban et~al.(2017)Serban, Sordoni, Lowe, Charlin, Pineau, Courville,
  and Bengio]{serban2017hierarchical-1}
Iulian~Vlad Serban, Alessandro Sordoni, Ryan Lowe, Laurent Charlin, Joelle
  Pineau, Aaron~C. Courville, and Yoshua Bengio.
\newblock A hierarchical latent variable encoder-decoder model for generating
  dialogues.
\newblock In \emph{AAAI}, pages 3295--3301, 2017.

\bibitem[Shang et~al.(2015)Shang, Lu, and Li]{shang2015neural}
Lifeng Shang, Zhengdong Lu, and Hang Li.
\newblock Neural responding machine for short-text conversation.
\newblock In \emph{ACL-IJCNLP}, pages 1577--1586, July 2015.

\bibitem[Shawar and Atwell(2007)]{shawar2007different}
Bayan~Abu Shawar and Eric Atwell.
\newblock Different measurements metrics to evaluate a chatbot system.
\newblock In \emph{Proceedings of the workshop on bridging the gap: Academic
  and industrial research in dialog technologies}, pages 89--96. Association
  for Computational Linguistics, 2007.

\bibitem[Shen et~al.(2014)Shen, He, Gao, Deng, and Mesnil]{shen2014latent}
Yelong Shen, Xiaodong He, Jianfeng Gao, Li~Deng, and Gr{\'e}goire Mesnil.
\newblock A latent semantic model with convolutional-pooling structure for
  information retrieval.
\newblock In \emph{Proceedings of the 23rd ACM International Conference on
  Conference on Information and Knowledge Management}, pages 101--110. ACM,
  2014.

\bibitem[Shum et~al.(2018)Shum, He, and Li]{shum2018xiaoice}
Heung{-}Yeung Shum, Xiaodong He, and Di~Li.
\newblock From eliza to xiaoice: Challenges and opportunities with social
  chatbots.
\newblock \emph{CoRR}, abs/1801.01957, 2018.
\newblock URL \url{http://arxiv.org/abs/1801.01957}.

\bibitem[Sordoni et~al.(2015)Sordoni, Galley, Auli, Brockett, Ji, Mitchell,
  Nie, Gao, and Dolan]{sordoni2015neural}
Alessandro Sordoni, Michel Galley, Michael Auli, Chris Brockett, Yangfeng Ji,
  Margaret Mitchell, Jian-Yun Nie, Jianfeng Gao, and Bill Dolan.
\newblock A neural network approach to context-sensitive generation of
  conversational responses.
\newblock In \emph{NAACL-HLT}, May 2015.

\bibitem[Sutskever et~al.(2014)Sutskever, Vinyals, and
  Le]{sutskever2014sequence}
Ilya Sutskever, Oriol Vinyals, and Quoc Le.
\newblock Sequence to sequence learning with neural networks.
\newblock In \emph{NIPS}, pages 3104--3112, 2014.

\bibitem[Sutton et~al.(1999)Sutton, Precup, and Singh]{sutton99between}
Richard~S. Sutton, Doina Precup, and Satinder~P. Singh.
\newblock Between {MDP}s and semi-{MDP}s: A framework for temporal abstraction
  in reinforcement learning.
\newblock \emph{Artificial Intelligence}, 112\penalty0 (1--2):\penalty0
  181--211, 1999.
\newblock An earlier version appeared as Technical Report 98-74, Department of
  Computer Science, University of Massachusetts, Amherst, MA 01003. April,
  1998.

\bibitem[Vedantam et~al.(2015)Vedantam, Lawrence~Zitnick, and
  Parikh]{vedantam2015cider}
Ramakrishna Vedantam, C~Lawrence~Zitnick, and Devi Parikh.
\newblock Cider: Consensus-based image description evaluation.
\newblock In \emph{Proceedings of the IEEE Conference on Computer Vision and
  Pattern Recognition}, pages 4566--4575, 2015.

\bibitem[Vinyals and Le(2015)]{vinyals2015neural}
Oriol Vinyals and Quoc Le.
\newblock A neural conversational model.
\newblock In \emph{ICML Deep Learning Workshop}, July 2015.

\bibitem[Vinyals et~al.(2015)Vinyals, Toshev, Bengio, and
  Erhan]{vinyals2015show}
Oriol Vinyals, Alexander Toshev, Samy Bengio, and Dumitru Erhan.
\newblock Show and tell: A neural image caption generator.
\newblock In \emph{Proceedings of the IEEE conference on computer vision and
  pattern recognition}, pages 3156--3164, 2015.

\bibitem[Wallace(2009)]{wallace2009anatomy}
Richard~S Wallace.
\newblock The anatomy of alice.
\newblock In \emph{Parsing the Turing Test}, pages 181--210. Springer, 2009.

\bibitem[Weizenbaum(1966)]{weizenbaum1966eliza}
Joseph Weizenbaum.
\newblock Eliza—a computer program for the study of natural language
  communication between man and machine.
\newblock \emph{Communications of the ACM}, 9\penalty0 (1):\penalty0 36--45,
  1966.

\bibitem[Wu et~al.(2016)Wu, Wang, and Xue]{wu2016ranking}
Bowen Wu, Baoxun Wang, and Hui Xue.
\newblock Ranking responses oriented to conversational relevance in chat-bots.
\newblock In \emph{Proceedings of COLING 2016, the 26th International
  Conference on Computational Linguistics: Technical Papers}, pages 652--662,
  2016.

\bibitem[Wu et~al.(2010)Wu, Burges, Svore, and Gao]{wu2010adapting}
Qiang Wu, Christopher~JC Burges, Krysta~M Svore, and Jianfeng Gao.
\newblock Adapting boosting for information retrieval measures.
\newblock \emph{Information Retrieval}, 13\penalty0 (3):\penalty0 254--270,
  2010.

\bibitem[Xing et~al.(2017)Xing, Wu, Wu, Liu, Huang, Zhou, and
  Ma]{xing2017topic}
Chen Xing, Wei Wu, Yu~Wu, Jie Liu, Yalou Huang, Ming Zhou, and Wei-Ying Ma.
\newblock Topic aware neural response generation.
\newblock In \emph{AAAI}, volume~17, pages 3351--3357, 2017.

\bibitem[Zhang et~al.(2016)Zhang, Wu, Wang, Zhou, and Li]{zhang2016learning}
Kai Zhang, Wei Wu, Fang Wang, Ming Zhou, and Zhoujun Li.
\newblock Learning distributed representations of data in community question
  answering for question retrieval.
\newblock In \emph{Proceedings of the Ninth ACM International Conference on Web
  Search and Data Mining}, pages 533--542. ACM, 2016.

\end{thebibliography}
